\pdfoutput=1
\documentclass[fleqn,usenatbib,useAMS]{mnras}

\usepackage{amssymb}	
\usepackage{multicol}        
\usepackage{bm}		
\usepackage{pdflscape}	
\usepackage[T1]{fontenc}
\usepackage{ae,aecompl}

\usepackage {graphicx}
\usepackage{natbib}
\usepackage{aas_journals}
\usepackage{color}
\usepackage{amsmath}
\usepackage{scalerel}

\usepackage{newtxtext,newtxmath}

\newcommand{\redmagic}{\textit{redMaGiC}}

\DeclareMathAlphabet{\mathbbm}{U}{bbm}{m}{n}
\newcommand{\mpc}{h^{-1}\mathrm{Mpc}}

\title[DES Y3 super-structures $\times$ {\it Planck} CMB lensing]{Dark Energy Survey Year 3 results: imprints of cosmic voids and superclusters in the {\it Planck} CMB lensing map}

\author[A. Kov\'acs et al. (the DES Collaboration)]{
\parbox{\textwidth}{
\Large
A.~Kov\'acs$^{1,2}$\thanks{Juan de la Cierva Fellow, corresponding author: \texttt{\rm \texttt{akovacs@iac.es}}}
P.~Vielzeuf$^{3}$,
I.~Ferrero$^{4}$,
P.~Fosalba$^{5,6}$,
U.~Demirbozan$^{7}$,
R.~Miquel$^{7,8}$,
C.~Chang$^{9,10}$,
N.~Hamaus$^{11}$,
G.~Pollina$^{11}$,
K.~Bechtol$^{12}$,
M.~Becker$^{13}$,
A.~Carnero~Rosell$^{1,2}$,
M.~Carrasco~Kind$^{14,15}$,
R.~Cawthon$^{16}$,
M.~Crocce$^{5,6}$,
A.~Drlica-Wagner$^{9,10,17}$,
J.~Elvin-Poole$^{18,19}$,
M.~Gatti$^{20}$,
G.~Giannini$^{7}$,
R.~A.~Gruendl$^{14,15}$,
A.~Porredon$^{18,19}$,
A.~J.~Ross$^{18}$,
E.~S.~Rykoff$^{21,22}$,
I.~Sevilla-Noarbe$^{23}$,
E.~Sheldon$^{24}$,
B.~Yanny$^{17}$,
T.~Abbott$^{25}$,
M.~Aguena$^{26,27}$,
S.~Allam$^{17}$,
J.~Annis$^{17}$,
D.~Bacon$^{28}$,
G.~Bernstein$^{20}$,
E.~Bertin$^{29,30}$,
S.~Bocquet$^{31}$,
D.~Brooks$^{32}$,
D.~Burke$^{21,22}$,
J.~Carretero$^{7}$,
F.~J.~Castander$^{5,6}$,
M.~Costanzi$^{33,34,35}$,
L.~N.~da Costa$^{26,36}$,
M.~E.~S. Pereira$^{37}$,
J.~De Vicente$^{23}$,
S.~Desai$^{38}$,
H.~T.~Diehl$^{17}$,
J.~Dietrich$^{39}$,
A.~Ferté$^{40}$,
B.~Flaugher$^{17}$,
J.~Frieman$^{10,17}$,
J.~Garcia-Bellido$^{41}$,
E.~Gazta\~{n}aga$^{5,6}$,
D.~Gerdes$^{37,42}$,
T.~Giannantonio$^{43,44}$,
D.~Gruen$^{31}$,
J.~Gschwend$^{26,36}$,
G.~Gutierrez$^{17}$,
S.~Hinton$^{45}$,
D.~L.~Hollowood$^{46}$,
K.~Honscheid$^{18,19}$,
D.~Huterer$^{37}$,
K.~Kuehn$^{47,48}$,
O.~Lahav$^{32}$,
M.~Lima$^{26,27}$,
M.~March$^{20}$,
J.~Marshall$^{49}$,
P.~Melchior$^{50}$,
F.~Menanteau$^{14,15}$,
R.~Morgan$^{12}$,
J.~Muir$^{51}$,
R.~Ogando$^{36}$,
A.~Palmese$^{52}$,
F.~Paz-Chinchon$^{14,43}$,
A.~Pieres$^{26,36}$,
A.~Plazas Malagón$^{50}$,
M.~Rodriguez Monroy$^{23}$,
A.~Roodman$^{21,22}$,
E.~Sanchez$^{23}$,
M.~Schubnell$^{37}$,
S.~Serrano$^{5,6}$,
M.~Smith$^{53}$,
E.~Suchyta$^{54}$,
G.~Tarle$^{37}$,
D.~Thomas$^{28}$,
C.-H.~To$^{18}$,
T.~N.~Varga$^{39,55,56}$,
J.~Weller$^{39,56}$
}
  \vspace{0.07cm}\\~\\
\parbox{\textwidth}{\centering \textsc{\Large(The DES Collaboration)} \\ \centering \textit{Author affiliations are listed at the end of this paper}\\ }}

\begin{document}
\label{firstpage}
\maketitle
\begin{abstract}
The CMB lensing signal from cosmic voids and superclusters probes the growth of structure in the low-redshift cosmic web. In this analysis, we cross-correlated the {\it Planck} CMB lensing map with voids detected in the Dark Energy Survey Year 3 (Y3) data set ($\sim$5,000 deg$^{2}$), expanding on previous measurements that used Y1 catalogues ($\sim$1,300 deg$^{2}$). Given the increased statistical power compared to Y1 data, we report a $6.6\sigma$ detection of negative CMB convergence ($\kappa$) imprints using approximately 3,600 voids detected from a \emph{redMaGiC} luminous red galaxy sample. However, the measured signal is \emph{lower} than expected from the MICE N-body simulation that is based on the $\Lambda$CDM model (parameters $\Omega_{\rm m} = 0.25$, $\sigma_8 = 0.8$), and the discrepancy is associated mostly with the void centre region. Considering the full void lensing profile, we fit an amplitude $A_{\kappa}=\kappa_{\scaleto{\rm DES}{4pt}}/\kappa_{\scaleto{\rm MICE}{4pt}}$ to a simulation-based template with fixed shape and found a moderate $2\sigma$ deviation in the signal with $A_{\kappa}\approx0.79\pm0.12$. We also examined the WebSky simulation that is based on a {\it Planck} 2018 $\Lambda$CDM cosmology, but the results were even less consistent given the slightly higher matter density fluctuations than in MICE.  We then identified superclusters in the DES and the MICE catalogues, and detected their imprints at the $8.4\sigma$ level; again with a lower-than-expected $A_{\kappa}=0.84\pm0.10$ amplitude. The combination of voids and superclusters yields a $10.3\sigma$ detection with an $A_{\kappa}=0.82\pm0.08$ constraint on the CMB lensing amplitude, thus the overall signal is $2.3\sigma$ weaker than expected from MICE.

\end{abstract}

\newpage

\section{Introduction}
\label{sec:Section1}

The largest under-dense structures in the low-redshift cosmic web, the \emph{cosmic voids}, are rich sources of cosmological information. The matter density profile, shape and redshift-dependent abundance of these cosmic voids all contain information about the physics of dark matter and the growth rate of structure \citep[see e.g.][]{sutter2014,pisani2015,Hamaus2016,Pollina2019,Verza2019,Hamaus2020,Nadathur2020,Aubert2020,Contarini2021}. In the context of dark energy, their imprints in the Cosmic Microwave Background (CMB) temperature map due to the Integrated Sachs-Wolfe (ISW) effect \citep{SachsWolfe} are also important complementary probes \citep[see e.g.][]{Granett2008,NadathurCrittenden2016,Cai2017,Kovacs2019} of dynamical nature, besides the more traditional geometrical probes of cosmic acceleration. Overall, voids have generated a growing interest when planning future galaxy survey projects with ever-increasing volumes, tracer density, and observational precision \citep[see e.g.][]{Pisani2019,Hamaus2021}. Complementing the statistical probes of over-dense environments, the expected gain in cosmological inference associated with voids comes from their ability to unravel extra information from the observed galaxy catalogues \citep[see e.g.][]{Kitaura2016,Nadathur2019,Paillas2021,Kreisch2021}. In particular, under-dense environments are prime candidates to detect differences between the standard and alternative cosmological models, and thus probe the nature of gravity \citep[see e.g.][]{Clampitt2013,Cai2015,Pollina2015,Cautun2018,Baker2018,Schuster2019,Davies2021}.

The gravitational \emph{lensing} signal from low-$z$ cosmic voids is also of great interest in observational cosmology, since it probes the relation of mass and light in a unique manner \citep[see e.g.][]{Fang2019}. Unlike clusters and filaments \citep[see e.g.][]{Baxter2015,Madhavacheril2015,He2017,Baxter2018}, cosmic voids cause a de-magnification effect and therefore correspond to local minima in the lensing convergence ($\kappa$) maps, estimated from the matter density field $\delta(r,\theta)$ via projection as
\begin{equation}
    \kappa(\theta)=\frac{3H_0^2\Omega_m}{2c^2}
    \int_{0}^{r_{\rm max}} \delta (r,\theta)
    \frac{(r_{\rm max}-r)r}{r_{\rm max}}\, dr,
    \label{eq:kappa_born}
\end{equation}
in the Born approximation with the Hubble constant $H_{\mathrm{0}}$ and matter density parameter $\Omega_{\mathrm{m}}$, where $r$ denotes the co-moving distance to source galaxies in the background of the lenses (with distorted shapes due to lensing), and $r_{\rm max}$ determines the maximum distance considered.

While the detection of the lensing signal, cosmic shear or convergence, from an individual void is challenging due to significant uncertainties \citep[][]{Amendola1999,Krause2013}, \emph{stacked} detections of the lensing signal from catalogues of voids have already been reported \citep{Melchior2014,Sanchez2016,Gruen2016,ClampittJain2015,Brouwer2018,Fang2019,Jeffrey2021}. To further optimise the signal-to-noise (S/N) of these void lensing measurements, N-body simulation analyses also support the observational work, including detailed probes of the signal characteristics given different void definitions \citep[see e.g.][]{Cautun2016,Davies2018,Davies2021}.

In recent years, another emerging field has been the analysis of the imprint of voids in reconstructed CMB lensing maps, primarily using \emph{Planck} data \citep[][]{Planck2018_cosmo}. Cross-correlations of void positions from the Baryon Oscillation Spectroscopic Survey (BOSS) data set and the \emph{Planck} $\kappa$ map first yielded a $3.2\sigma$ detection of void lensing effects \citep[][]{Cai2017}. A better understanding of different sub-classes of voids \citep[see e.g.][]{NadathurEtal2016} and the application of an advanced matched filter detection technique \citep[][]{NadathurCrittenden2016} then helped to increase the significance of the CMB $\kappa$ signal imprinted by BOSS voids from the $3.2\sigma$ \citep[][]{Cai2017} level to $5.3\sigma$ \citep[][]{Raghunathan2019}.

Based on spectroscopic datasets, the cross-correlation results from BOSS voids $\times$ \emph{Planck} CMB lensing were generally consistent with expectations from the standard $\Lambda$-Cold Dark Matter ($\Lambda$CDM) model. However, more recent voids $\times$ CMB lensing measurements using photometric galaxy redshifts from the Dark Energy Survey \citep[DES,][]{DES_review} Year 1 (Y1) data set (1,300 deg$^{2}$) by \cite{Vielzeuf2019} reported a lensing amplitude parameter $A_{\kappa}=\kappa_{\scaleto{\rm DES}{4pt}}/\kappa_{\scaleto{\rm MICE}{4pt}}\approx0.72\pm0.17$, implying about $20\%$ lower-than-expected signals. However, this finding depended slightly on the specific void population and analysis method considered, and the observed deviations from the expected $A_{\kappa}\approx1$ were not highly significant and thus no important anomaly was reported. A similar CMB lensing analysis by \cite{Hang2021} also showed approximately $20\%$ weaker signals than expected, using a combination of voids and superclusters detected from the Dark Energy Spectroscopic Instrument \citep[DESI,][]{DESI} Legacy \emph{Imaging} Survey \citep[DESI-LS,][]{Dey2019} photo-$z$ data set (17,700 deg$^{2}$). We note that, while these DES Y1 and DESI-LS findings look generally consistent, differences in cosmological parameters in the simulation analyses and different void/supercluster finding strategies make the detailed comparisons challenging.

In a wider context, we note that the lensing and ISW effects probe two different, yet physically related, properties of the gravitational potential; the lensing effects probe \emph{spatial variations} of the potential, while the ISW effect is sensitive to its \emph{time-dependence}. It is interesting to consider that possible connections may exist between these moderate CMB lensing tensions and observed anomalies in the ISW imprints of voids. The ISW amplitude parameter ($A_{\scaleto{\rm ISW}{4pt}}$ = $\Delta T_\mathrm{obs}/\Delta T_{\scaleto{\rm \Lambda CDM}{4pt}}$, defined as the ratio of the observed and the expected $\Delta T_{\scaleto{\rm ISW}{4pt}}$ signals) has often been found significantly higher from $R\gtrsim100~\mpc$ voids, or supervoids, than expected in the concordance model \citep{Granett2008,Cai2017,Kovacs2016,Kovacs2018}, including these main findings:

\begin{itemize}

\item Combining BOSS and DES Year 3 data using similar supervoid definitions \citep[][]{Kovacs2019}, an excess ISW amplitude with $A_{\scaleto{\rm ISW}{4pt}}\approx5.2\pm1.6$ was reported from about 200 supervoids in the $0.2<z<0.9$ redshift range ($A_{\scaleto{\rm ISW}{4pt}}\approx1$ in the $\Lambda$CDM model). 

\item Testing the extremes, a recent mapping of the Eridanus supervoid aligned with the CMB Cold Spot using DES weak lensing \emph{mass maps} \citep[][]{Jeffrey2021} also provides more evidence for the alignment of a large under-dense region and a colder-than-expected spot on the CMB sky \citep[][]{Kovacs2022}.

\item Curiously, no significant excess ISW signals were reported when using matched filters and different void definitions in BOSS analyses \citep[][]{NadathurCrittenden2016}, or when using an alternative DESI-LS galaxy catalogue that approximately maps the same BOSS+DES footprint in the same redshift range, considering a slightly different void definition \citep[][]{Hang2021}. 

\item The enhanced ISW signals from supervoids are also considered anomalous because 2-point correlation analyses do \emph{not} show significant excess either, compared to $\Lambda$CDM predictions \citep[see e.g.][]{PlanckISW2015,Stolzner2018,Hang20212pt}.

\end{itemize}

In this paper, we identify cosmic voids and superclusters from the DES Year 3 data set (5,000 deg$^{2}$) to better understand the observed moderate tensions concerning the ISW and lensing imprints of these structures in the \emph{Planck} CMB maps. The paper is organised as follows. In Section \ref{sec:Section2}, we introduce our observed and simulated data sets and our stacking methodology. We then present our main observational results in Section \ref{sec:Section3}. Then, Section \ref{sec:Section4} contains a discussion, and in Section  \ref{sec:Section5} we present our main conclusions.

\section{Data sets $\&$ Methods}
\label{sec:Section2}
\subsection{Dark Energy Survey Year-3 data}

\begin{figure}
\begin{center}
\includegraphics[width=88mm
]{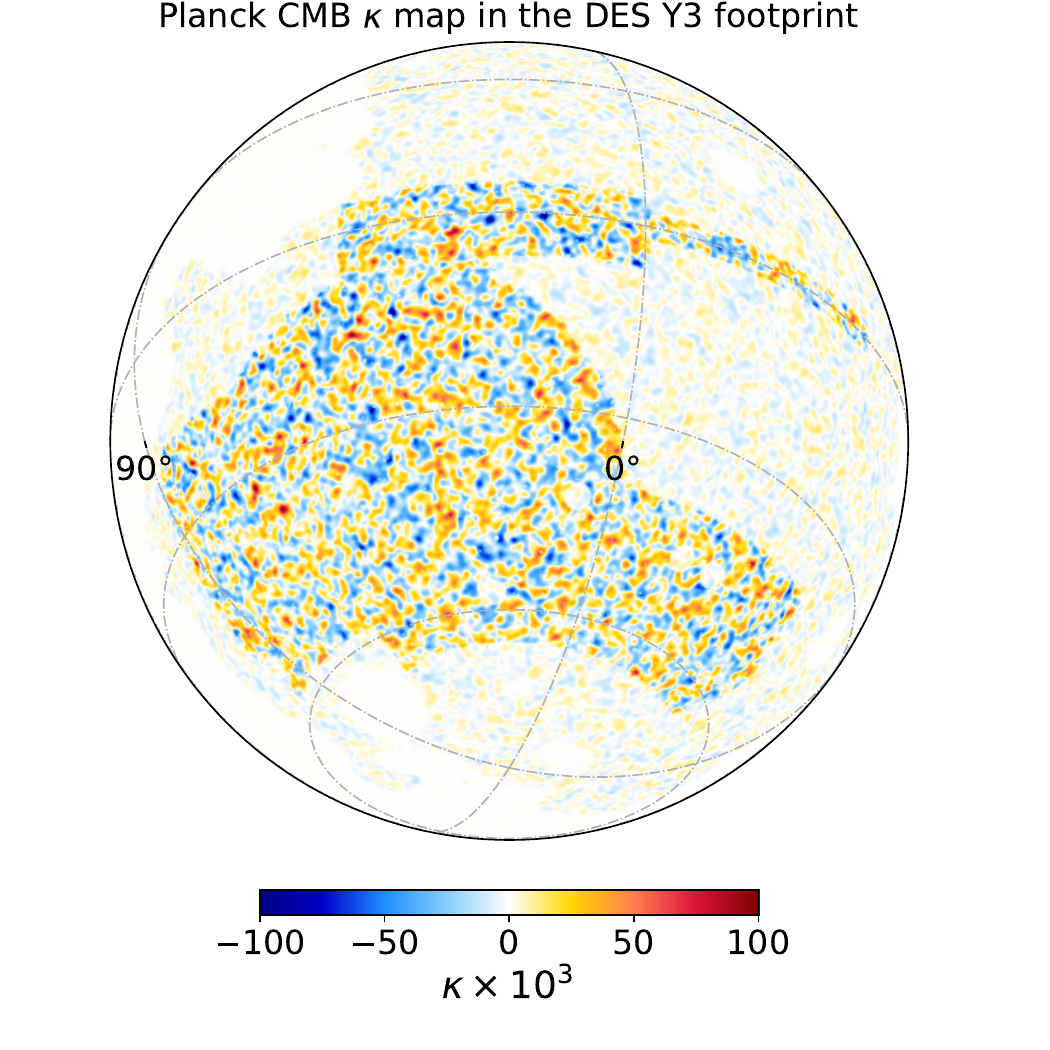}
\end{center}
\caption{The Year 3 footprint of the Dark Energy Survey is highlighted on the \emph{Planck} CMB lensing convergence map in Equatorial coordinates. We applied a FWHM=$1^{\circ}$ Gaussian smoothing to suppress noise contributions from small-scale fluctuations.}
\label{fig:figure_area}
\end{figure}

We mapped the cosmic web using the DES Year 3 (Y3) data set. DES is a six-year survey that covers approximately $5000~\mathrm{\deg}^2$ sky area of the South Galactic Cap (see Figure \ref{fig:figure_area}). Mounted on the Cerro Tololo Inter-American Observatory (CTIO) four metre Blanco telescope in Chile, the $570$~megapixel Dark Energy Camera \citep[DECam,][]{Flaugher2015} images the field in $grizY$ filters. The raw images were processed by the DES Data Management (DESDM) team  \citep{Sevilla2011,Morganson2018}. We adopted the empirically constructed DES Y3 survey mask in our analysis, which excludes potentially contaminated pixels e.g. in the close proximity of bright stars. We note that the Y3 data set already covers the full DES survey footprint, and the final Y6 data set will provide a deeper imaging over the same area. For the full details of the DES Y3 data set, we refer the readers to \cite{y3-gold}.

To identify voids, we used a catalogue of luminous red galaxies (LRG), photometrically selected by the red-sequence MAtched-filter Galaxy Catalog \cite[\redmagic,][]{Rozo2015} method, that is based on the red-sequence MAtched-filter Probabilistic Percolation (redMaPPer) cluster-finding algorithm \citep{Rykoff2014}. The \redmagic\ sample spans the $0.15<z<0.8$ range, and its principal advantage in the void-finding context is the exquisite $\sigma_\mathrm{z}/(1+z)\approx 0.02$, photo-$z$ precision and a low outlier rate.

The resulting DES Y3 LRG sample has an approximately constant co-moving space density of galaxies with $\bar{n}\approx 4\times10^{-4}h^{3}$ $\mathrm{Mpc^{-3}}$ (high-luminosity sample, brighter than 1.0$L_{*}$). While the full 3D position information of the LRGs is not accessible using the DES \redmagic\ photo-$z$ data, the great photo-$z$ precision allows a robust reconstruction of the largest cosmic voids. We highlight that the \redmagic\ galaxy samples have successfully been used in a series of DES void analyses including weak lensing and ISW measurements \citep[see e.g.][]{Gruen2016,Sanchez2016,Kovacs2016, Fang2019, Vielzeuf2019}. Further details about the general galaxy clustering properties of the latest DES Y3 \redmagic\ data set are presented by \cite{Pandey2021}.

\subsection{Simulations of galaxy catalogues and $\kappa$ maps}

We simulated our DES Y3 lensing measurements using the MICE (Marenostrum Institut de Ciencias de l'Espai) N-body simulation \citep[][]{Fosalba2015,MICE3,Carretero2015}, which spans $(3 h^{-1}\mathrm{Gpc})^3$ co-moving volume based on the \texttt{GADGET2} code \citep{springel2005}. It assumes a $\Lambda$CDM model with input cosmological parameters $\Omega_m=0.25$, $\Omega_\Lambda=0.75$, $\Omega_b=0.044$, $n_s=0.95$, $\sigma_8=0.8$ and $h=0.7$ from the Five-Year Wilkinson Microwave Anisotropy Probe (WMAP) results \citep{Komatsu2009}.

The MICE CMB $\kappa$ map was created based on projected 2D pixel maps of the convergence field using the "Onion Universe" approach \citep[][]{Fosalba2008}, proven to be successful in producing maps with lensing power spectra in agreement with the Born
and Limber approximations \citep[see e.g.][]{Jain2000}. The dark-matter is added up in “onion shells”, or projected density maps, in the MICE light-cone, weighted by the weak-lensing efficiency at each redshift.
The map is provided with a pixel resolution of $N_{\rm side}=2048$ \citep[see][for more details]{Fosalba2015}. However, given the nature of our problem and the relatively large degree-scale angular size of voids, we downgraded the high resolution map to a lower $N_{\rm side}=512$ resolution. The downgraded map matches the resolution of the {\it Planck} $\kappa$ map that we used in this analysis.

We then created a DES-\emph{like}
MICE {\redmagic} light-cone mock catalogue based on a Halo Occupation Distribution (HOD) methodology \citep[see e.g.][]{Tinker2012}. While the detailed description of this process is beyond the scope of this paper, we direct the interested readers to two related DES HOD analyses. Our mock construction is based on the findings by \cite{Zacharegkas2021}, who fitted an HOD model to the galaxy-galaxy lensing signal and the number density of the DES Y3 {\redmagic} sample in wide redshift bins. It was found that {\redmagic} galaxies typically live in dark matter halos of mass $\log_{10}(M_{h}/M_{\odot}) \approx 13.7$, without significant dependence on redshift. In our implementation, we closely followed the mock construction methodology\footnote{An HOD modelling approach developed for a DES galaxy sample optimised for reconstructions of the baryonic acoustic oscillation (BAO) signal.} by \cite{Ferrero2021}, and carried out a similar HOD analysis in the MICE light-cone. Our mock reproduced the $n(z)$ redshift distribution of LRGs observed in the DES Y3 data, leading also to consistent void and supercluster samples.

We note that the MICE cosmology is relatively far from the best-fit {\it Planck} cosmology \citep{Planck2018_cosmo} that is often used as a reference. For instance, the two cosmologies differ in the values of $\Omega_m$ and $H_0$, and such differences are expected to affect the overall amplitude of the lensing signal (see Eq. \ref{eq:kappa_born}). Another influential factor is the $\sigma_8$ parameter for determining the matter content \citep[][]{Nadathur2019} and the lensing convergence \citep[see e.g.][]{Davies2021} of voids; its value in the MICE simulation ($\sigma_8=0.8$) is quite close to the best-fit {\it Planck} 2018 value ($\sigma_8=0.811\pm0.006$). Consequently, the CMB $\kappa$ signal in MICE is expected to be weaker than from a {\it Planck} cosmology, due to an approximate $C_{\ell}^{\rm g\kappa}\sim\Omega_{\rm m}^{0.78}\sigma_8$ scaling with the most relevant cosmological parameters of the basic $\Lambda$CDM model, as determined by \cite{Hang20212pt} considering a similar redshift range (there is also an estimated weaker $A_{\kappa}\sim h^{0.24}$ scaling with the Hubble constant). Furthermore, the linear galaxy bias parameter $b$ also acts as an overall re-scaling factor for the amplitude of the galaxy-CMB lensing cross-correlation signal. Thus any imperfection in its modelling may also lead to differences between simulations and observed data.

As a further consistency test of our methodology, we also analysed the publicly available\footnote{https://mocks.cita.utoronto.ca/data/websky/} \emph{WebSky} simulation \citep{websky1,websky2} to model the stacked CMB lensing signal of voids. Importantly, the \emph{WebSky} mock is based on the {\it Planck} 2018 cosmology \citep{Planck2018_cosmo}. It provides a light-cone halo catalogue and, among other data products, a corresponding CMB lensing $\kappa$ map that we used in our analyses.

\subsection{Void and supercluster finding}
\label{sec:void_finder}

While numerous algorithms exist to define cosmic voids, we used the so-called 2D void finding algorithm which was developed to deal with photo-$z$ data sets, where a full 3D information is inaccessible due to smearing effects in the line-of-sight \citep{Sanchez2016}. The 2D void definitions have shown great potential to extract void lensing signals in N-body simulations \citep[see e.g.][]{Cautun2018}, with slightly better performance than using 3D methods including empty spheres techniques \citep[see e.g.][]{Hawken2017,Zhao2016}. The 2D voids are typically larger in radius than typical 3D voids, and the associated mean lensing potential fluctuations are also larger. However, \cite{Fang2019} verified that 3D void definitions, such as the widely used \texttt{ZOBOV} method \citep[][]{ZOBOV}, can also be successfully applied to detect void lensing signals. We note that voids selected from photo-$z$ data sets are on average preferentially elongated in the line-of-sight \citep[see e.g.][]{Kovacs2016}, and thus correspond to high $S/N$ lensing and ISW detections, using either 2D or 3D definitions \citep[see e.g.][]{Fang2019}.

To capitalise on this counter-intuitive advantage associated with using photo-$z$ data, large samples of 2D voids have been used in previous DES void lensing and ISW measurements, showing robust detections from both observed data and simulations \citep[see e.g.][]{Vielzeuf2019,Kovacs2019}. 

\begin{figure*}
\begin{center}
\includegraphics[width=170mm
]{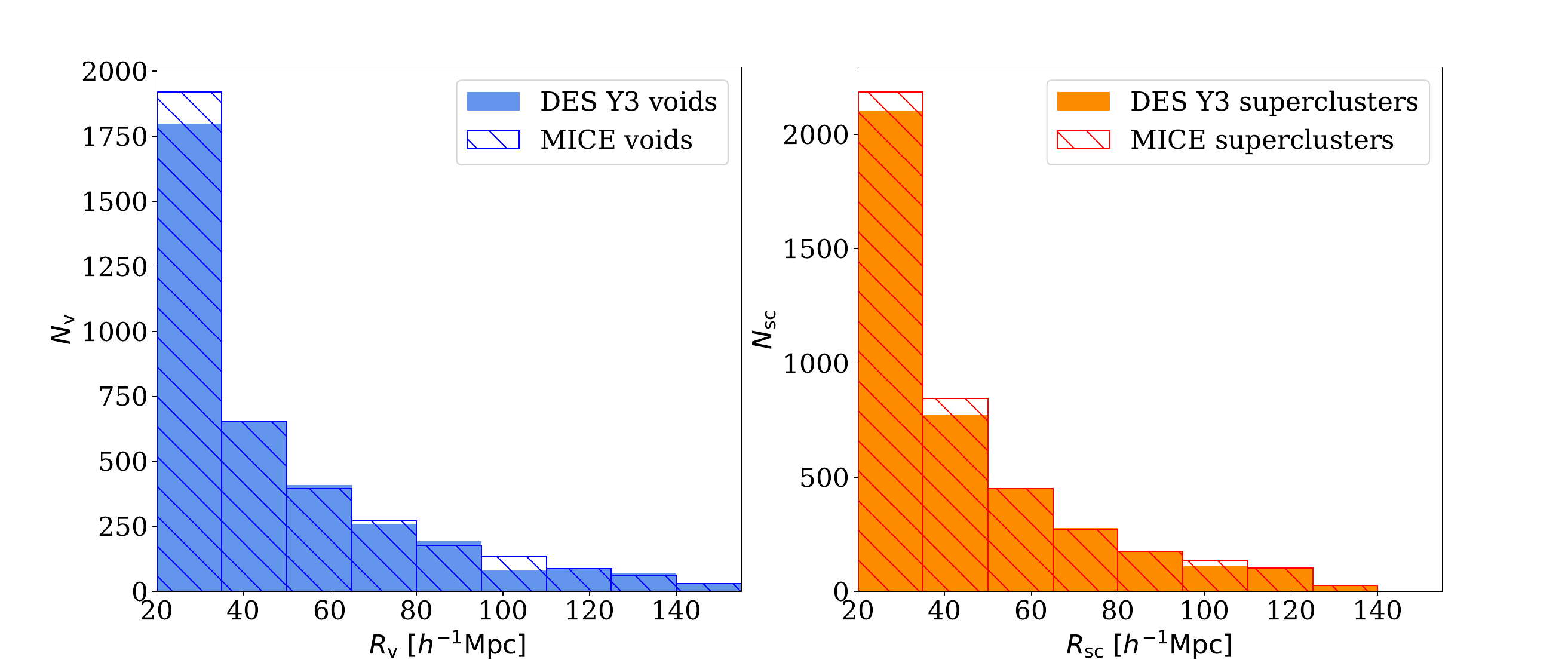}
\end{center}
\caption{Comparison of void (left) and supercluster (right) radius distributions extracted from MICE and DES Y3. Overall, we report a great agreement between simulations and observations for both voids and superclusters. We applied a minimum radius cut $R\gtrsim20~\mpc$ to eliminate potentially spurious under-densitites and over-densitites, while the largest void and supercluster radii are approximately $R\approx150~\mpc$ in both simulated and observed catalogues.} 
\label{fig:figure_Nz}
\end{figure*}

The 2D void-finding method is based on tomographic slices of galaxy data, and an analysis of the projected density field around void centre candidates, defined by minima in the smoothed density field. 
The void radius $R_\mathrm{v}$ is defined using an algorithm that includes measurements of galaxy density in annuli about void centres until the mean density is reached \citep[see][for details]{Sanchez2016}. Practically measured in degrees, $R_\mathrm{v}$ is then converted to units of $\mpc$ assuming the best-fit {\it Planck} cosmology \citep{Planck2018_cosmo}. We note that our main analyses of the void lensing profiles are based on stacked images in units of re-scaled relative void radii $R/R_\mathrm{v}$, and thus the actual unit of the void radius is not highly important. We note that there are three main free parameters in this process:
\begin{itemize}
    \item \emph{Thickness of the tomographic slices:} it was determined that an $s\approx100~\mpc$ line-of-sight slicing (roughly consistent with the typical photo-$z$ scatter of LRGs) effectively leads to the detection of independent, and individually significant under-densities \citep{Sanchez2016,Kovacs2019}. We thus sliced the DES Y3 galaxy catalogue into 15 shells of $100~\mpc$ thickness at $0.15<z<0.8$. 
    \item \emph{Density map smoothing scale:} another void-finding parameter is the Gaussian smoothing scale applied to the tracer density map to define density minima. To facilitate lensing analyses, it was shown in previous DES analyses \citep[see e.g.][]{Sanchez2016,Vielzeuf2019} that a $\sigma=10~\mpc$ smoothing of the maps allows a robust detection of voids which carry most of the lensing signal. We therefore followed \cite{Vielzeuf2019} and used $\sigma=10~\mpc$ as a galaxy density map smoothing parameter.
\item \emph{Central pixel density:} a third parameter is the minimum central under-density of a pixel in smoothed density maps that is considered as a void centre. We again followed \cite{Kovacs2019} and \cite{Vielzeuf2019} and selected voids with at least $30\%$ under-density in their centres, which ensures that too shallow and thus potentially spurious voids are excluded. 
\end{itemize}

For exploratory CMB lensing cross-correlation analyses, we also identified \emph{superclusters} using a similar 2D methodology. Following \cite{Kovacs2016}, we inverted our 2D void finder to identify over-densities in the smoothed galaxy density field. We then radially measured the projected galaxy density in 2D maps around such significant positive peaks until the mean density is reached, as a definition of the supercluster radius ($R_\mathrm{sc}$).

In Figure \ref{fig:figure_Nz}, we show histograms of the void and supercluster radii for observed DES data and from the MICE simulations. The typical radius of the voids and superclusters we identified is about $R\approx50~\mpc$, while the maximum void and supercluster radius is approximately $R\approx150~\mpc$. On the other end of the distribution, we pruned the void and supercluster catalogues by applying a $R\gtrsim20~\mpc$ cut to eliminate potentially spurious objects due to photo-$z$ scatter.

Using the DES Y3-\emph{like} mock catalogue selected from MICE with the HOD methodology, the observed and simulated voids and superclusters show great agreement in their radius distributions, which is an important validation of our methods (see Figure \ref{fig:figure_Nz}).

\begin{figure*}
\begin{center}
\includegraphics[width=81mm]{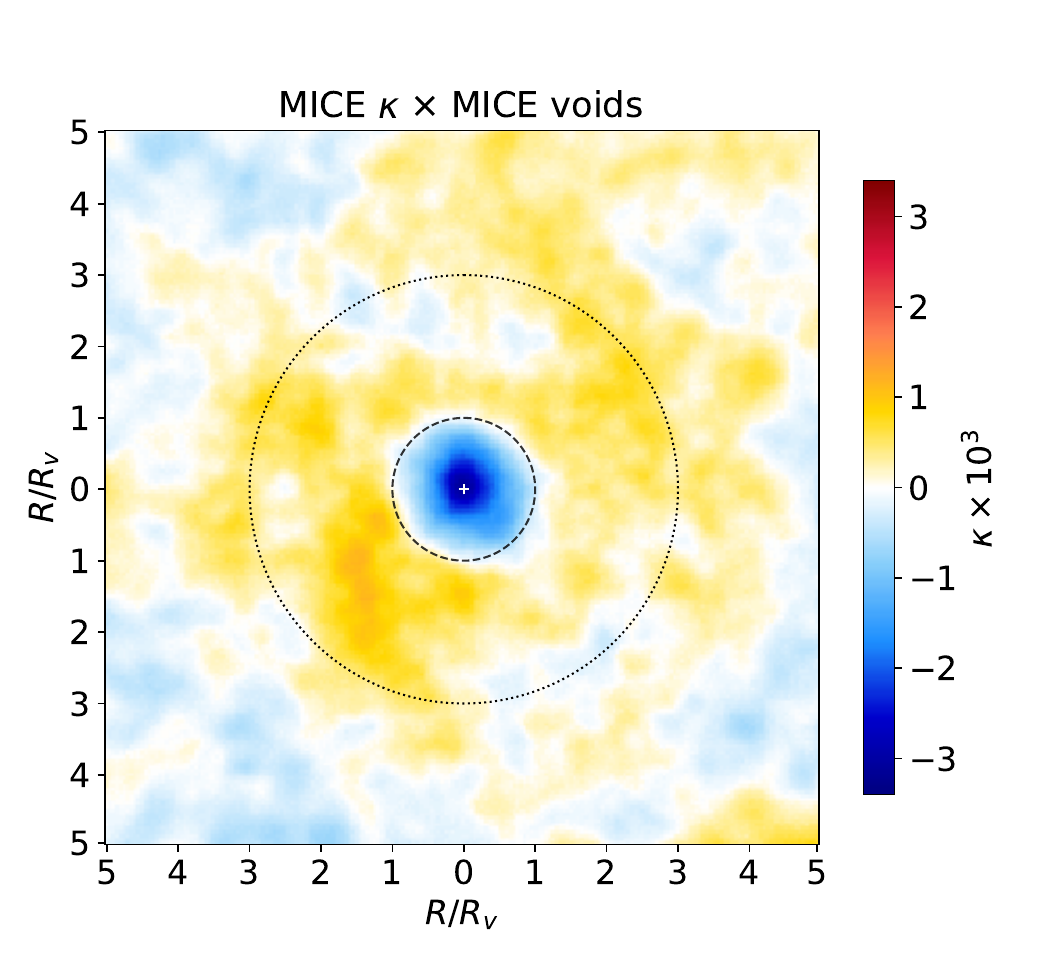}
\includegraphics[width=81mm]{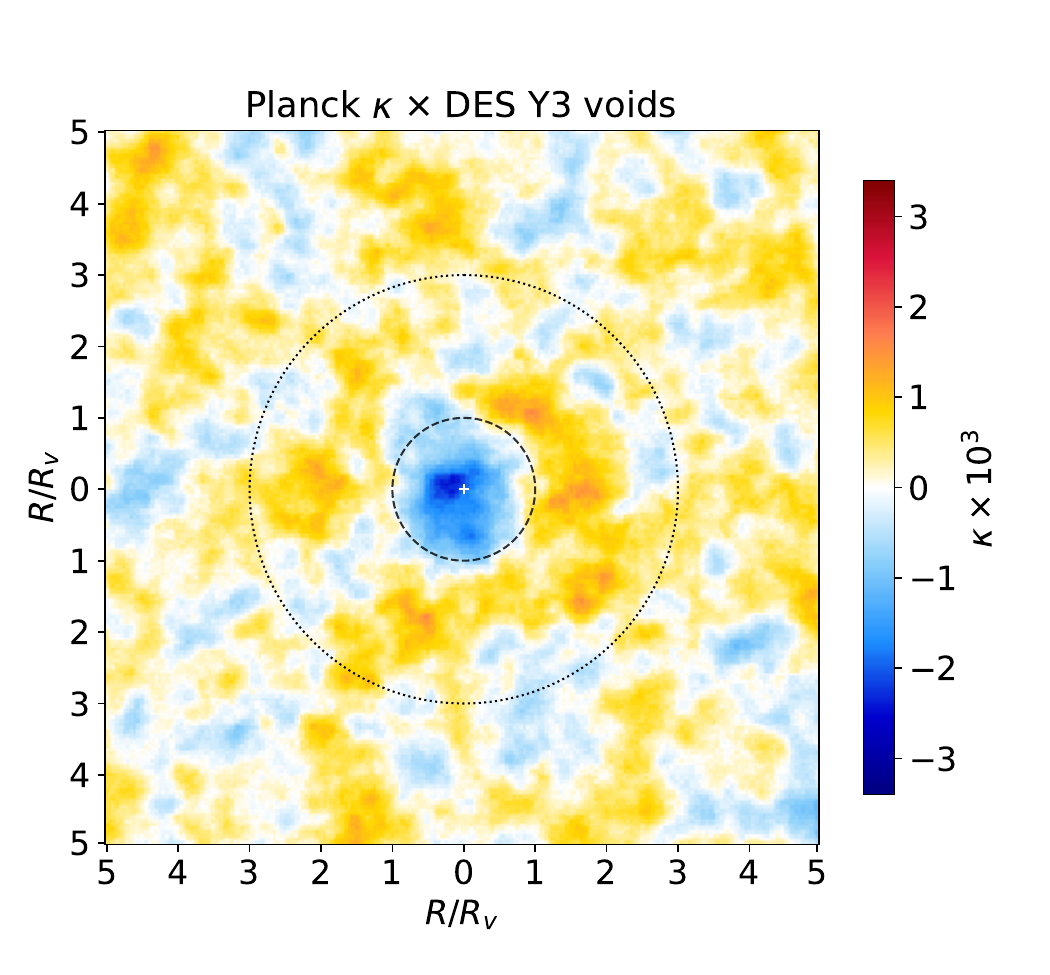}\\
\includegraphics[width=81mm]{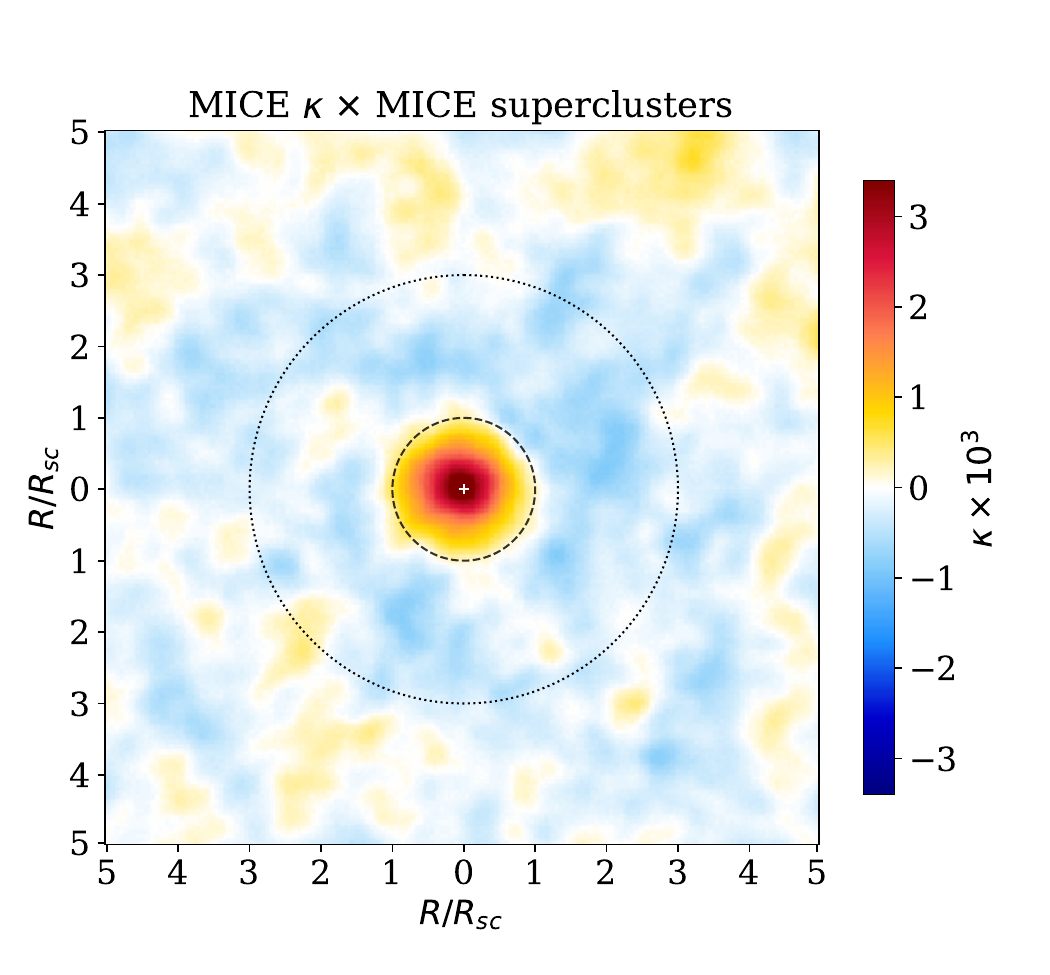}
\includegraphics[width=81mm]{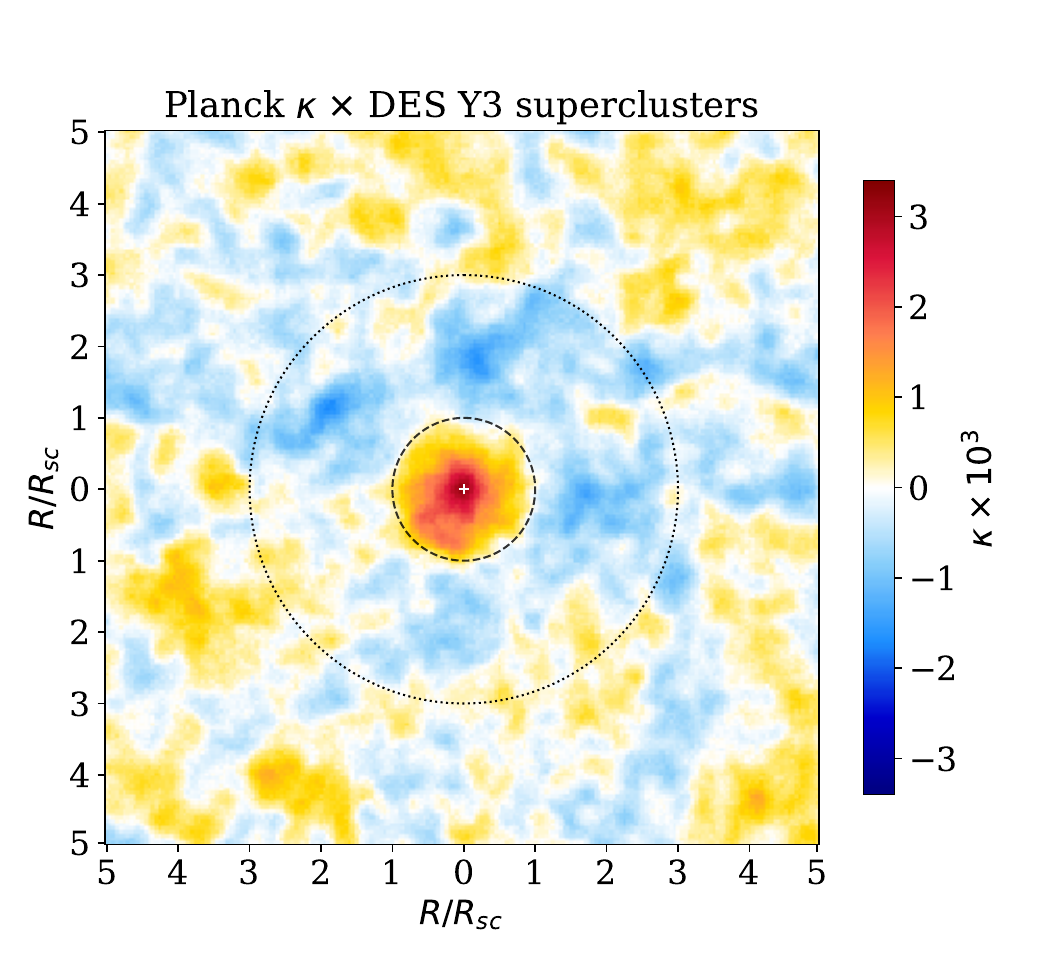}
\end{center}
\caption{The stacked CMB $\kappa$ signal from voids (top panels) and superclusters (bottom panels) in MICE (left) vs. DES Y3 (right). The dashed (dotted) circles mark one (three) void and supercluster radius in re-scaled units. The observed data displays a lower lensing signal for both voids and superclusters, especially visible in the centre. The MICE images are based on a \emph{noiseless} simulation to better characterise the true signal.}
\label{fig:stacked_images}
\end{figure*}

\subsection{CMB lensing map from the {\it Planck} data set}

In our measurements, we used the reconstructed Minimum Variance CMB $\kappa$ map provided by the {\it Planck} collaboration, released in the form of $\kappa_{\rm lm}$ spherical harmonic coefficients \citep[see][for details]{Planck2018_lensing} up to  $\ell_{\rm max}=2048$. We created a $\kappa$ map at $N_{\rm side}=512$ resolution by converting the $\kappa_{\rm lm}$ values into \texttt{healpix} maps \citep{Healpix}, and also constructed a corresponding mask from the publicly available {\it Planck} data. We also tested our pipeline using the previous (2015) version of the {\it Planck} CMB lensing map and found consistent results.

We note that even though higher resolution maps may be extracted from the $\kappa_{\rm \ell m}$ coefficients, $N_{\rm side}=512$ is a sufficient choice given the degree-size angular scales involved in our problem. In order to further suppress small-scale noise effects, we applied a FWHM=$1^{\circ}$ Gaussian smoothing to the convergence map, following the S/N optimisation efforts by \cite{Vielzeuf2019}. This smoothing was consistently applied to observed and simulated $\kappa$ maps.

\subsection{Stacking measurement method}

Following a standard methodology in the field \citep[see for instance][]{Vielzeuf2019,Hang2021}, we \emph{stacked} cut-out patches of the CMB $\kappa$ maps to measure the mean imprint of voids, since the lensing imprint of individual voids is too weak to be observed with high confidence due to the significant noise levels. 

Given the angular void size that is determined by the void finder for each object, we first cut out square-shaped patches from the CMB $\kappa$ maps aligned with void positions using the \texttt{gnomview} projection method of the \texttt{healpix} package \citep{Healpix}. In this process, we \emph{re-scaled} the size of the cut-out images knowing the void radius $R_\mathrm{v}$. 

The re-scaling method guarantees that void centres, void edges, and the compensation walls in their surroundings are all in alignment for different voids using the chosen $R/R_\mathrm{v}$ units \citep[see e.g.][for alternative techniques without re-scaling]{Raghunathan2019}. In order to examine the compensation region around the void interiors ($R/R_\mathrm{v}<1$), we selected a wider area up to five void radii ($R/R_\mathrm{v}<5$) around each void. To estimate the mean imprint of voids in the CMB $\kappa$ map. We then measured tangential $\kappa$ profiles from the stacked images using 16 bins of $\Delta (R/R_\mathrm{v})=0.3$ width. 

\begin{figure*}
\begin{center}
\includegraphics[width=180mm
]{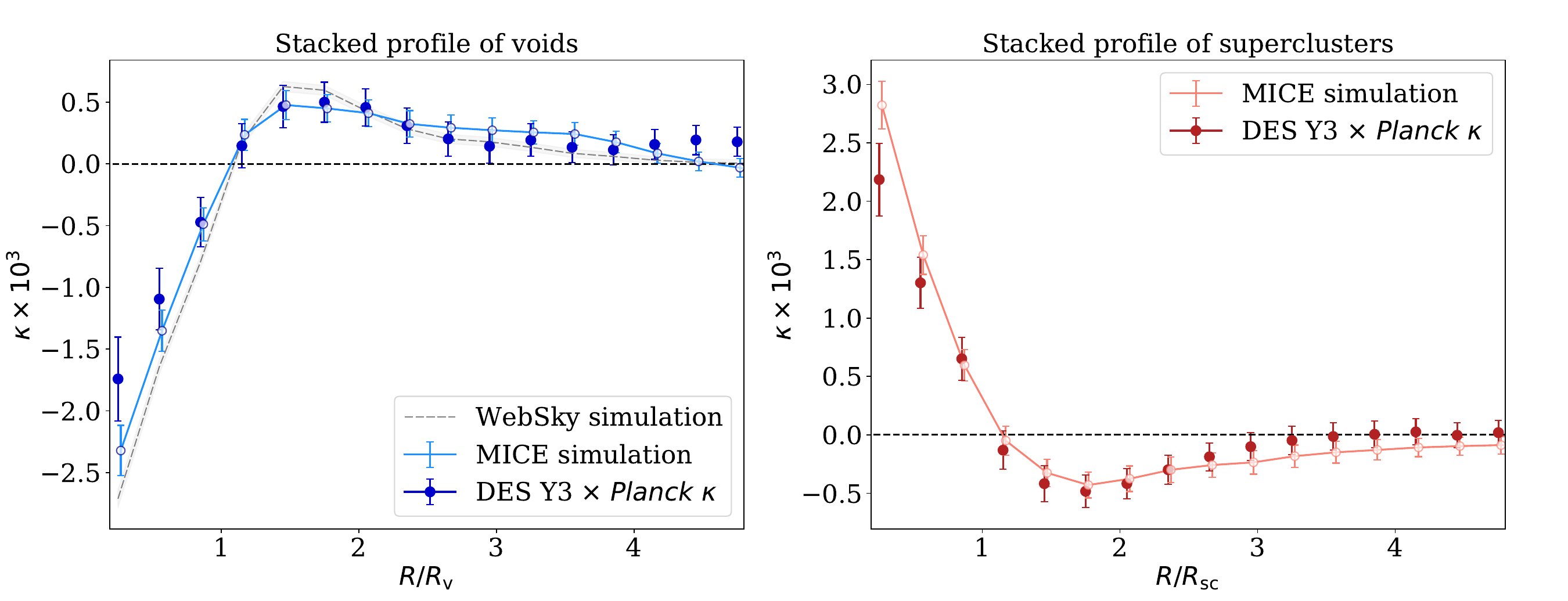}
\end{center}
\caption{Radial profiles measured from the stacked CMB $\kappa$ images are shown for voids (left) and superclusters (right). We detected a lower-than-expected signal from the DES Y3 data in both cases. The disagreement compared to the WebSky simulation's \emph{Planck} cosmology is even greater than compared to MICE.}
\label{fig:profiles}
\end{figure*}

\section{Results}
\label{sec:Section3}

\subsection{Expected lensing signals from MICE}

To characterise the expected signal given the $\Lambda$CDM cosmology implemented in MICE, we first identified 3,727 voids in the MICE {\redmagic}-like catalogue (approximately 5,000 deg$^{2}$ sky area in one octant) using the 2D void definition and selection criteria that we described in Section \ref{sec:void_finder}. We also identified 4,195 superclusters in the MICE mock data set by inverting our 2D void-finding algorithm. 

We then stacked the MICE CMB $\kappa$ map on the positions of these mock voids and superclusters. We note that, even without observational effects, the estimation of the mean signal of simulated voids and superclusters from the projected CMB $\kappa$ map has an \emph{intrinsic} scatter. Since these structures are not isolated objects, their complicated overlap distorts the $\kappa$ signal generated by their gravitational potential, which only allows a statistical measurement rather than an idealistic determination of their individual imprints. To determine this scatter, we did not add a realistic observational noise to the signal-only MICE $\kappa$ map in our first simulation analyses. However, in the calculation of actual observational errors, additional fluctuations from a {\it Planck}-like measurement noise in the CMB $\kappa$ maps (as well as the role of the DES Y3 survey mask) are taken into account and they mostly add small-scale fluctuations to the stacked images of voids and superclusters.

We note that, while the simulated MICE $\kappa$ map is provided in a full-sky format, the mock \redmagic\ galaxy catalogue only spans one octant and therefore we applied a simple octant mask to the lensing map. We also transformed the $\kappa$ map to a zero-mean version by subtracting the mean $\bar{\kappa}\approx10^{-4}$ value in the octant we analysed.

The main results from these simulated stacking measurements are presented in the left panels of Figure \ref{fig:stacked_images}. We detected a clear negative (positive) CMB $\kappa$ imprint from voids (superclusters) in the MICE mock catalogue in the central region and up to the re-scaled radius of the structures ($R/R_{v}<1$). In contrast, we found a slightly positive (negative) $\kappa$ signal in the surroundings of voids (superclusters) from the compensating over-dense (under-dense) regions at $1<R/R_{v}<5$ (see Figure \ref{fig:stacked_images}). Overall, these finding are consistent with expectations and also DES Y1 results by \cite{Vielzeuf2019} who analysed smaller patches (1,300 deg$^{2}$) in the MICE simulation using the same void definition and stacking methodology.

\subsection{DES Y3 measurements and covariance matrix}

Following the methodology developed using the MICE simulation, we detected 3,578 voids and 4,010 superclusters from DES Y3 data, which also covers about $5000~\mathrm{\deg}^2$. We found a clear detection of negative (positive) CMB $\kappa$ signals aligned with the centres of voids (superclusters), as shown in the right panels of Figure \ref{fig:stacked_images}. While these stacked images using the \emph{Planck} CMB $\kappa$ map naturally feature larger noise fluctuations compared to the stacked images from MICE without observational noise (see the left panels of Figure \ref{fig:stacked_images}), we nonetheless detected a lower amplitude in the lensing signal from DES Y3 voids and superclusters in the centre of the images.

The dominant source of the stacking measurement uncertainties is the random instrumental noise in the {\it Planck} data. Additionally, the total error also has contributions from uncertainties in the CMB $\kappa$ signal generated by the voids, with about half the magnitude of the instrumental noise (see Figure \ref{fig:profiles} for further details). This second uncertainty is, at least in part, due to the fact that the mean $\kappa$ imprint of voids is not calculated as an average of several \emph{isolated} structures. Instead, the complicated overlap structure of the voids and superclusters along the line-of-sight, and the overlap with the surroundings of their neighbour structures in the same redshift slice inevitably results in an imperfect reconstruction of the imprints from these structures, starting from the CMB $\kappa$ map's fluctuations.

To account for these two sources of uncertainty in the cross-correlations, we created random realisations of the signal-only CMB $\kappa$ maps and also random noise maps using the noise power spectrum released by the {\it Planck} team \citep{Planck2018_lensing}. Following \cite{Vielzeuf2019}, we measured the power spectrum of the signal-only MICE $\kappa$ map $S_{\rm \kappa}^{\scaleto{\rm MICE}{4pt}}$ using the \texttt{anafast} routine of \texttt{healpix}. Then, given the same power spectrum, we created 1000 random $\kappa$ map realisations using \texttt{synfast}. To model the noise, we also generated 1000 {\it Planck}-like noise map realisations. We then added our $N_{\rm \kappa}^{i}$ noise maps to the $S_{\rm \kappa}^{i}$ MICE-like CMB $\kappa$ map realisations. Finally, we stacked these 1000 random $S_{\rm \kappa}^{i}+N_{\rm \kappa}^{i}$ convergence maps on the uncorrelated DES Y3 void and supercluster positions to characterise the standard deviation of chance fluctuations in such cross-correlation measurements. While the DES Y3 and MICE catalogues of voids and superclusters are in good agreement, we decided to use the observed DES Y3 catalogues for the estimation of the errors to ensure that the mask effects, overlap structure, or any other correlation between voids and superclusters is fully realistic.

We applied a further $10\%$ correction to take into account that we use a single realisation of voids and superclusters instead of varying their positions when cross-correlated with random $\kappa$ maps (determined by \cite{cabre_etal} using simulations). Therefore, we re-scaled our covariance matrix and increased the errors on the measurement of the CMB lensing amplitude by about $10\%$ for a conservative analysis which incorporates this additional variance (see \cite{Kovacs2019} for a similar DES void $\times$ CMB analysis with such a correction applied).

\subsection{Template-fitting analysis of radial profiles}

To quantify the visual impressions presented in Figure \ref{fig:stacked_images}, we then measured CMB $\kappa$ profiles from the stacked images. We used re-scaled radius units using 16 bins of $\Delta (R/R_\mathrm{v})=0.3$ width up to five times the void radius ($R/R_\mathrm{v}=5$). This is the summary statistic that we used to compare the lensing imprint of voids in the MICE simulation and in DES Y3 data; together with the random cross-correlation measurements which characterise the covariance.

We present our main results in Figure \ref{fig:profiles} where the observed CMB $\kappa$ signals from the DES Y3 data are compared to the corresponding MICE-based expectations. While the zero-crossing of the lensing signal is accurately recovered (at $R/R_{v}\approx1$), we found that the DES Y3 results are about $\sim20\%$ lower than expected from MICE analyses in the centres of both the voids and the superclusters. As listed in Table \ref{tab_significances}, this corresponds to a mild $1.8\sigma$ tension.

A complicating aspect is the possible difference between the $\Lambda$CDM $\kappa$ template profiles extracted from simulations based on slightly different cosmological parameters. Following \cite{Vielzeuf2019}, we also compared the DES Y3 results to the estimated CMB $\kappa$ imprints from the WebSky simulation that is, unlike MICE, based on a \emph{Planck} 2018 cosmology  \citep[see][for details]{websky1,websky2}. In this exploratory analysis, we applied a simple halo mass cut with $M>10^{13.5}h^{-1}M_\odot$ to define an LRG-like population which models the DES Y3 \redmagic\ sample. We also added Gaussian photo-$z$ errors with a $\sigma_z/(1+z)\approx 0.02$ scatter to the simulated WebSky spec-$z$ coordinates to create realistic observational conditions. As shown in Figure \ref{fig:profiles}, the WebSky results feature a slightly stronger void lensing profile than the MICE simulation, i.e. it is even less consistent with the DES Y3 results. While field-to-field fluctuations are non-negligible and there are differences in the simulated analyses, the MICE-WebSky comparison suggests that a more comprehensive analysis with different cosmological parameters might help to better understand these moderate tensions, and determine how exactly the void lensing signal depends on cosmology.

To measure the consistency between DES Y3 and MICE, we constrained the best-fitting $A_{\kappa}=\kappa_{\scaleto{\rm DES}{4pt}}/\kappa_{\scaleto{\rm MICE}{4pt}}$ amplitude parameter (and its error $\sigma_{\rm A_{\kappa}}$) as a ratio of the observed and simulated CMB $\kappa$ signals using the full radial profile. We again followed the DES Y1 analysis by \cite{Vielzeuf2019} and evaluated the statistic
\begin{equation}
{\chi}^2 = \sum_{ij} \left(\kappa_{i}^{\scaleto{\rm DES}{4pt}}-A_{\rm \kappa}\cdot\kappa_{i}^{\scaleto{\rm MICE}{4pt}}\right) C_{ij}^{-1} \left(\kappa_{j}^{\scaleto{\rm DES}{4pt}}-A_{\rm \kappa}\cdot\kappa_{j}^{\scaleto{\rm MICE}{4pt}}\right)
\end{equation}
where $\kappa_{i}$ is the mean CMB lensing signal in radius bin $i$, and $C$ is the corresponding covariance matrix. We searched for a best-fitting $A_{\rm \kappa}\pm\sigma_{\rm A_{\kappa}}$ amplitude by \emph{fixing} the shape of the stacked convergence profile to that calibrated from the MICE simulation.

We note that, while informative to better understand the data, this effective 1-parameter $A_{\rm \kappa}$ fit to the expected profile shape introduces a form of model-dependence to our analysis. Even though we can expect a good agreement between the simulations and the DES Y3 observations based on the Y1 results by \cite{Vielzeuf2019}, more complicated deviations may emerge in the real-world data which are hard to capture in detail with this statistic. We nonetheless expressed our constraints in this format to match the standards of the field, making our DES Y3 findings more easily comparable to results in the literature.

As detailed above, we estimated the covariance using 1000 randomly generated $\kappa$ maps with MICE-like power spectrum and \emph{Planck}-like noise. We also corrected our estimates with an Anderson-Hartlap factor $\alpha=(N_{\rm randoms}-N_{\rm bins}-2)/(N_{\rm randoms}-1)$, providing a $\approx2\%$ correction given our DES Y3 measurement setup \citep{hartlap2007}.

\begin{table}
\centering
\caption{\label{tab_significances} We compare $A_{\rm \kappa}$ constraints using all the voids and superclusters to analyses using sub-sets of the catalogues. We found a trend for a lower-than-expected signal coming mostly from deeper voids. We also observed that the low-$z$ half of the sample and larger voids with $R_{\rm v}>35~\mpc$ show weaker signals that the other half.}
\begin{tabular}{@{}ccccc}
\hline
\hline
{\bf Voids} & $N_{\rm v}^{\scaleto{\rm DES}{4pt}}$ & $A_{\rm \kappa}\pm\sigma_{A_{\rm \kappa}}$ & S/N & Tension \\
\hline
all objects & 3578 & $0.79\pm0.12$ & 6.6 & $1.8\sigma$ \\
$0.15<z<0.55$ & 1600 & $0.55\pm0.23$ & 2.4 & $2.0\sigma$ \\
$0.55<z<0.8$ & 1978 & $0.88\pm0.13$ & 6.8 & $0.9\sigma$ \\
$R_{\rm v}<35~\mpc$ & 1799 & $0.82\pm0.16$ & 5.1 & $1.1\sigma$ \\
$R_{\rm v}>35~\mpc$ & 1779 & $0.66\pm0.15$ & 4.4 & $2.3\sigma$ \\
$\delta_{\rm c}<-0.6$ & 2031 & $0.56\pm0.14$ & 4.0 & $3.1\sigma$ \\
$\delta_{\rm c}>-0.6$ & 1547 & $0.95\pm0.20$ & 4.8 & $0.3\sigma$ 
\\
\hline
\hline
{\bf Superclusters} & $N_{\rm sc}^{\scaleto{\rm DES}{4pt}}$ & $A_{\rm \kappa}\pm\sigma_{A_{\rm \kappa}}$ & S/N & Tension \\
\hline
all objects & 4010 & $0.84\pm0.10$ & 8.4 & $1.6\sigma$ \\
$0.15<z<0.55$ & 1942 & $0.70\pm0.15$ & 4.7 & $2.0\sigma$ \\
$0.55<z<0.8$ & 2068 & $0.91\pm0.13$ & 7.0 & $0.7\sigma$ \\
$R_{\rm sc}<35~\mpc$ & 2103 & $0.91\pm0.14$& 6.5 & $0.6\sigma$ \\
$R_{\rm sc}>35~\mpc$ & 1907 & $0.75\pm0.14$ & 5.4 & $1.8\sigma$ \\
$\delta_{\rm c}>0.9$ & 2102 & $0.89\pm0.13$ & 6.8 & $0.8\sigma$ \\
$\delta_{\rm c}<0.9$ & 1908 & $0.83\pm0.15$ & 5.5 & $1.1\sigma$ 

\\
\hline
{\bf Combined} & 7588 & $0.82\pm0.08$ & 10.3 & $2.3\sigma$
\\
\hline
\hline
\end{tabular}
\end{table}

\begin{figure}
\begin{center}
\includegraphics[width=82mm
]{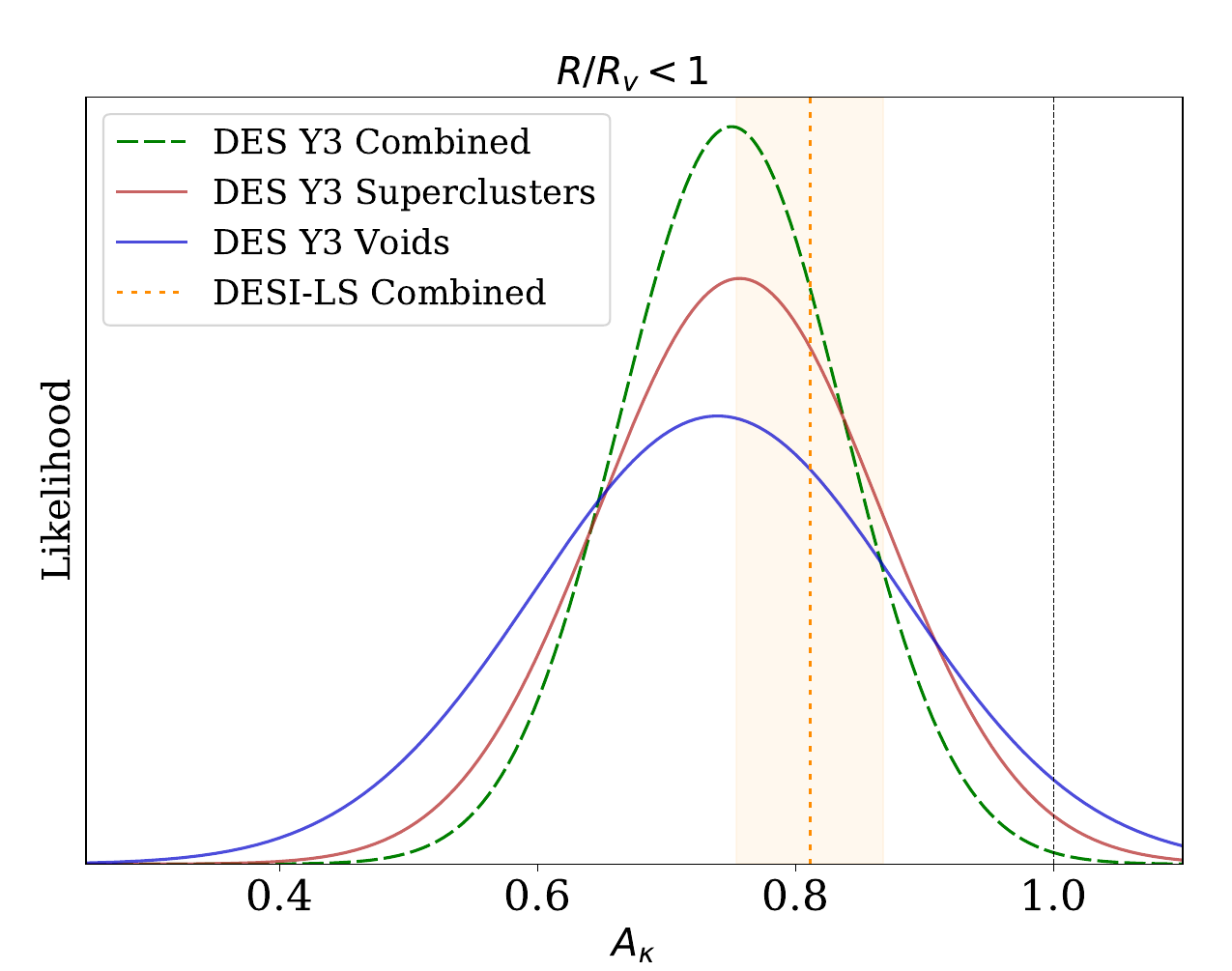}\\
\includegraphics[width=82mm
]{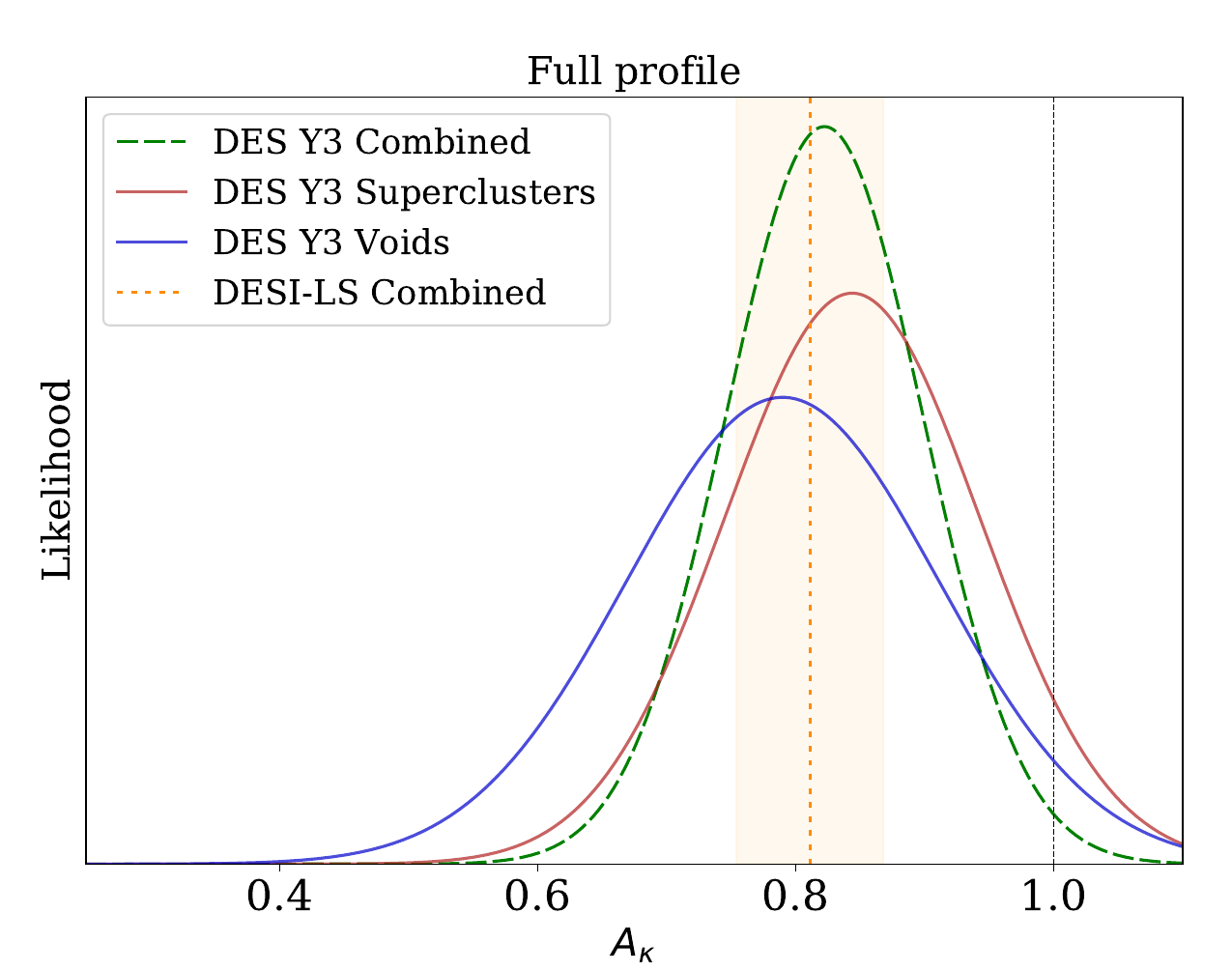}
\end{center}
\caption{Top: constraints on the $A_{\rm \kappa}$ lensing amplitude parameter using the first 3 radial bins ($R/R_{v}<1$) in the DES Y3 void (blue) and supercluster (red) interiors. Bottom panel: constraints on $A_{\rm \kappa}$ from all the 16 radial bins up to 5 times the void radius. We compare our findings, including combined results (green), to recent constraints from the DESI Legacy \emph{Imaging} Survey.}
\label{fig:profile_like}
\end{figure}

In particular, we aimed to test the hypothesis that the DES Y3 and MICE results are in close agreement which would imply $A_{\rm \kappa}\approx1$. At the same time, our statistical tests also reveal the \emph{detection significance} compared to zero signal, i.e. $A_{\rm \kappa}=0$, that is independent from the assumed model for the CMB $\kappa$ profile amplitude.

Given a MICE-like signal and the uncertainties from our DES Y3 $\times$ \emph{Planck} setup, we estimated that the $A_{\rm \kappa}$ parameter can be measured with approximately $10\%$ precision compared to a fiducial $A_{\rm \kappa}^{\rm fid}$ amplitude, both for voids and superclusters, which is equivalent to a $S/N\approx10$ detection: $A_{\rm \kappa}/A_{\rm \kappa}^{\rm fid}\approx1.0\pm0.12$ for the voids, and $A_{\rm \kappa}/A_{\rm \kappa}^{\rm fid}\approx1.0\pm0.10$ for the superclusters, and $A_{\rm \kappa}/A_{\rm \kappa}^{\rm fid}\approx1.0\pm0.08$ for their combination. As expected, this precision is higher than the result from DES Y1 (1,300 deg$^{2}$) with $A_{\rm \kappa}/A_{\rm \kappa}^{\rm fid}\approx1.0\pm0.17$ by \cite{Vielzeuf2019}, but it is slightly less accurate than the state-of-the-art $A_{\rm \kappa}/A_{\rm \kappa}^{\rm fid}\approx1.0\pm0.05$ constraint detectable from the DESI-LS (17,700 deg$^{2}$) measurement setup \citep{Hang2021}.

\subsection{Main results on the CMB lensing amplitude}

From the MICE template-fitting analysis, we determined that the overall best-fit CMB lensing amplitude from DES Y3 voids is $A_{\rm \kappa}\approx0.79\pm0.12$, equivalent to $S/N\approx6.6$ significance but about $1.8\sigma$ lower than expected from MICE. If the analysis is restricted to the first 3 radial bins ($R/R_{v}<1$), then the DES data is consistent with $A_{\rm \kappa}\approx0.74\pm0.14$ which further confirms our observation that the discrepancy comes predominantly from the void centre regions. 

Using the DES Y3 superclusters, we found $A_{\rm \kappa}\approx0.84\pm0.10$ ($S/N\approx8.4$ detection, $1.6\sigma$ tension with MICE), that is consistent with the results from the voids. Restricted to the central $R/R_{v}<1$ part, the best-fitting amplitude is $A_{\rm \kappa}\approx0.76\pm0.11$ that is again consistent with the corresponding result for voids. 

Considering the full profiles, the combination of voids and superclusters yields an $A_{\rm \kappa}\approx0.82\pm0.08$ constraint on the CMB lensing amplitude (with an assumption that void and supercluster signals are independent), i.e. a significant $S/N\approx10.3$ detection and an overall $2.3\sigma$ tension with MICE. These results are explained in Figure \ref{fig:profile_like}, in comparison with the $A_{\rm \kappa}\approx0.811\pm0.057$ constraint from a combination of voids and superclusters in the DESI-LS data set. We note that \cite{Hang2021} reported $A_{\rm \kappa}\approx0.937\pm0.087$ for voids alone, and $A_{\rm \kappa}\approx0.712\pm0.076$ for superclusters, which add up to the combined DESI-LS result.

\begin{figure*}
\begin{center}
\includegraphics[width=170mm
]{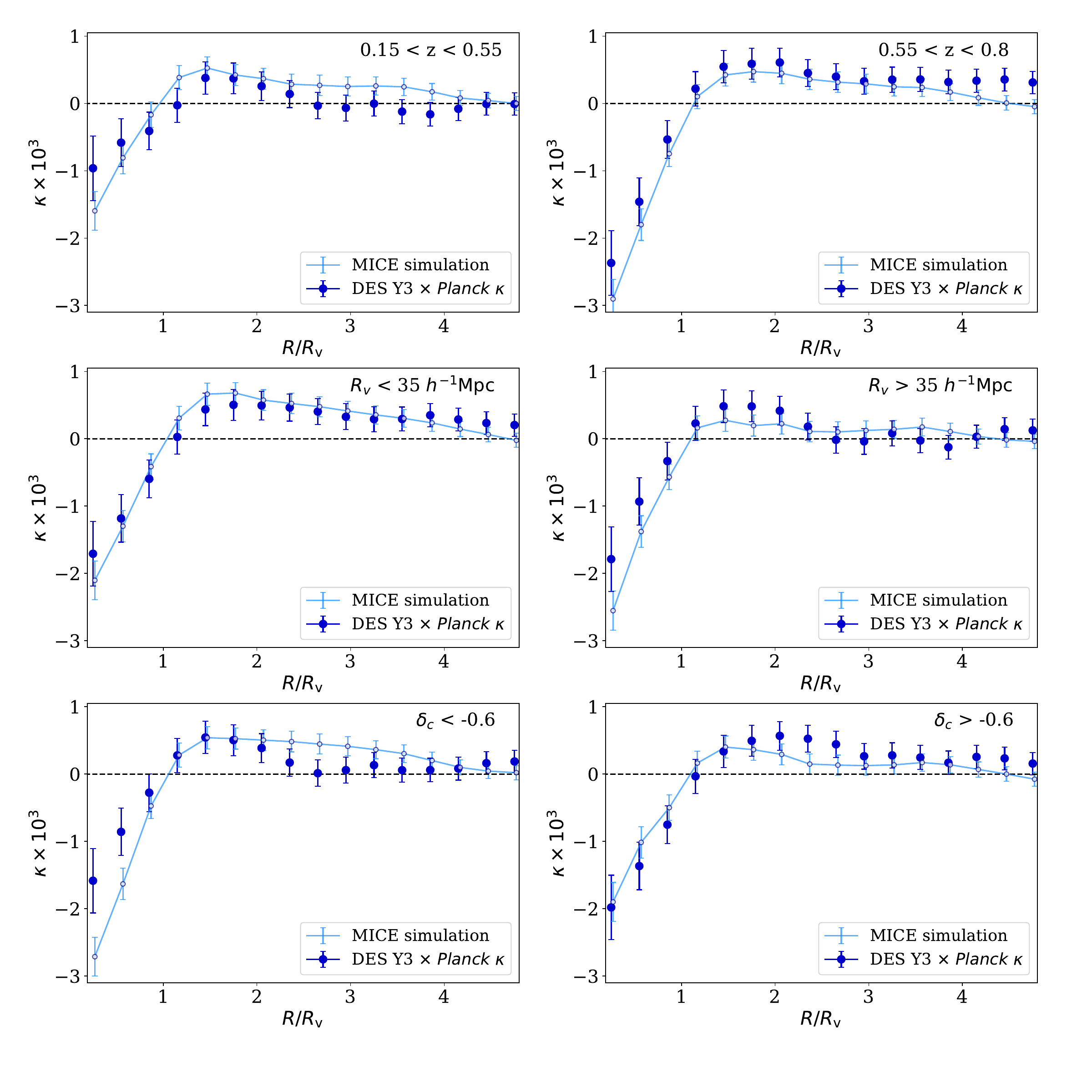}
\end{center}
\caption{We split the DES Y3 and MICE void samples into roughly two equal parts based on redshift (top), radius (middle), and under-density (bottom) to test possible trends in the lower-than-expected CMB lensing signal. We found that the lower redshift part of the data prefers lower lensing amplitudes, while larger and deeper voids also show more discrepancy compared to MICE. Further details about the numerical best-fit amplitude values are available in Table \ref{tab_significances}.}
\label{fig:profile_voidsub}
\end{figure*}

\begin{figure*}
\begin{center}
\includegraphics[width=170mm
]{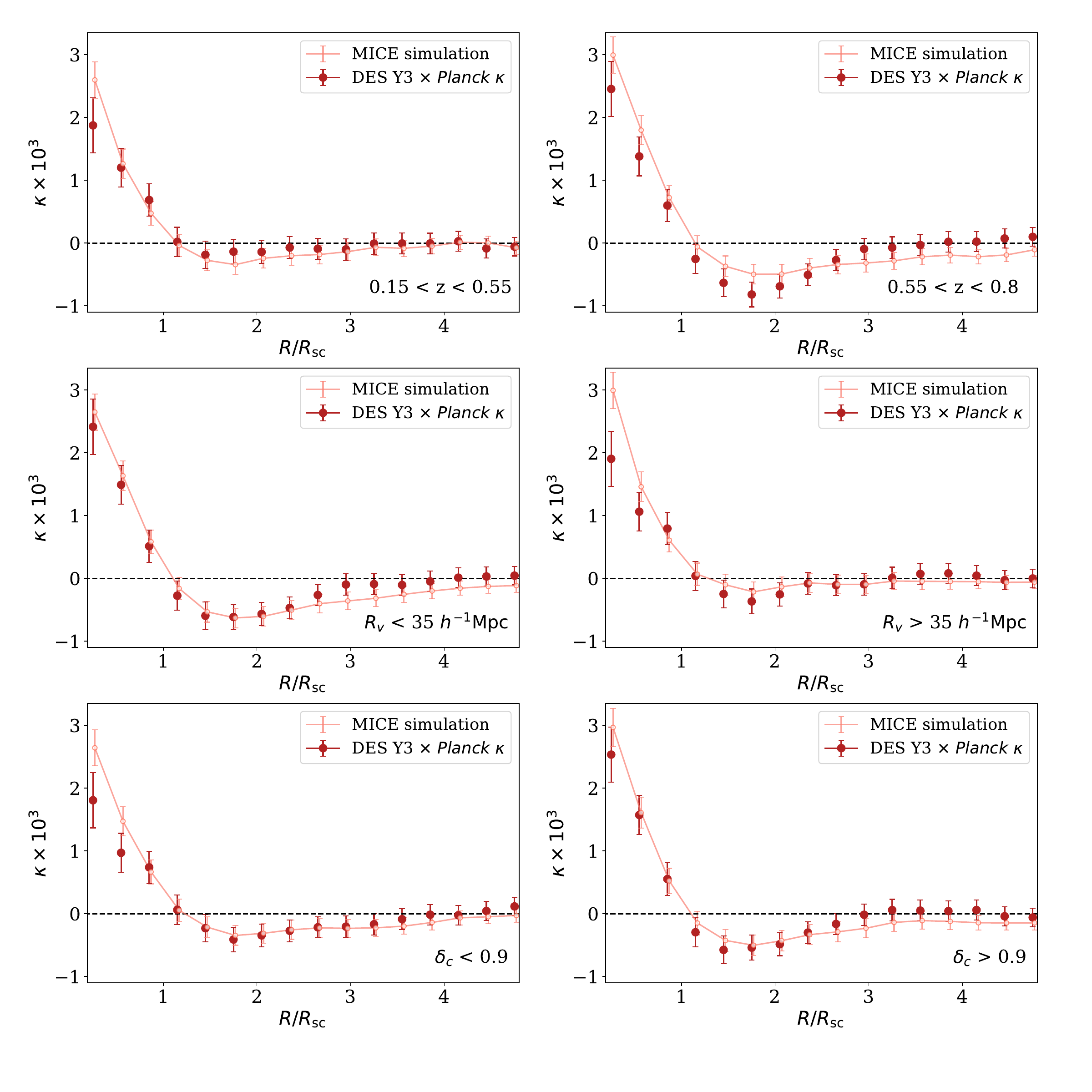}
\end{center}
\caption{We split the DES Y3 and MICE supercluster samples into roughly two equal parts based on redshift (top), radius (middle), and under-density (bottom) to test possible trends in the lower-than-expected CMB lensing signal. Compared to voids, we found better internal consistency between bins but noted that low-$z$ superclusters prefer lower lensing amplitudes. Moreover, we observed that shallower and larger superclusters also show more discrepancy, especially in their central regions. Further details about the numerical best-fit amplitude values are available in Table \ref{tab_significances}.}
\label{fig:profile_scsub}
\end{figure*}

Looking for the origins of the lower-than-expected $\kappa$ imprints, we split the DES Y3 void and supercluster catalogues roughly in half based on a number of different properties, and measured the stacked $\kappa$ imprints in the resulting subsets. The results of these exploratory analyses (considering the full lensing profile) are presented in Figure \ref{fig:profile_likelihoods_comp}. In Table \ref{tab_significances}, we also present further details about the number of voids and superclusters in each of the subsets, and also list the corresponding best-fit $A_{\rm \kappa}$ values. We made the following main observations:
\begin{itemize}
\item Voids at $0.15<z<0.55$ showed a trend for lower $\kappa$ signals with $A_{\rm \kappa}\approx0.55\pm0.23$, i.e. equivalent to a moderate $2.0\sigma$ tension with the MICE expectations. This lower-than-expected signal exists throughout most of the lensing profile including the void centre and the compensation zone (see Figure \ref{fig:profile_voidsub}).
\item Larger voids with $R_{\rm v}>35~\mpc$ show evidence for a weakened signal in the centre with best-fit amplitude $A_{\rm \kappa}\approx0.66\pm0.15$. This moderate $2.3\sigma$ tension is also visible in Figure \ref{fig:profile_voidsub}.
\item The deeper half of the DES Y3 voids with $\delta_{\rm c}<-0.6$ featured a $3.1\sigma$ tension with a low $A_{\rm \kappa}\approx0.56\pm0.14$ amplitude. As shown in Figure \ref{fig:profile_voidsub}, this lower-than-expected signal mostly comes from the void interiors ($R/R_\mathrm{v}<1$), but a reduced signal is also seen in the outer compensation zone at $2<R/R_\mathrm{v}<4$.
\item While generally being more consistent than voids, superclusters at $0.15<z<0.55$, and also the larger half of the objects with $R_{\rm v}>35~\mpc$, both showed a trend for a lower-than-expected signal at $R/R_\mathrm{v}<1$ compared to the other halves of the sample. We present further details in Figure \ref{fig:profile_scsub}. 
\item We also observed a weak trend for lower $\kappa$ signals from shallower superclusters ($\delta_{\rm c}<0.9$) with $A_{\rm \kappa}\approx0.83\pm0.15$, but these outcomes are consistent with the results from the complete sample.
\item Overall, we found that imprints of voids and superclusters at higher redshift (and lower density contrast) are quite symmetric to the zero line in absolute value (comparing the $\kappa$ signal amplitudes in the centres of voids and superclusters). However, a hint of asymmetry (higher absolute values of $\kappa$ in superclusters than in voids, when comparing top-left panels in Figures  \ref{fig:profile_voidsub} and  \ref{fig:profile_scsub}) is observed at low redshifts, as a possible manifestation of the rising non-linearity.
\end{itemize}

We note that further analyses of how the void definition and details in the stacking methodology affect the results are presented in a companion paper \citep[][]{Demirbozan2022}, which generally confirms the main findings of this analysis.

\section{Discussion}
\label{sec:Section4}

In this paper, we investigated the imprints of cosmic voids and superclusters in the CMB lensing convergence ($\kappa$) maps reconstructed from \emph{Planck} data. We used the Year 3 data set from the Dark Energy Survey to detect these large-scale structures in the cosmic web traced by \redmagic\ LRGs (see Figures \ref{fig:figure_area} and \ref{fig:figure_Nz}), and measured their stacked CMB lensing signals (see Figure \ref{fig:stacked_images}) which can constrain the low-$z$ growth of structure in a way that is complementary to typical large-scale structure probes. 
We then compared the DES Y3 observations to expectations from the MICE simulation \citep[][]{Fosalba2015}. We determined that DES Y3 voids feature a highly significant ($S/N\approx6.6$), albeit lower-than-expected, signal with $A_{\kappa}=\kappa_{\scaleto{\rm DES}{4pt}}/\kappa_{\scaleto{\rm MICE}{4pt}}=0.79\pm0.12$ amplitude (see Figures \ref{fig:profiles} and \ref{fig:profile_like} for details), in comparison with the $\Lambda$CDM model implemented in the MICE simulation  (parameters $\Omega_{\rm m} = 0.25$, $\sigma_8 = 0.8$). Analysing DES Y3 superclusters, our detection significance reached the $8.4\sigma$ level, but again with a lower-than-expected $A_{\kappa}\approx0.84\pm0.10$ amplitude. We found that these moderate $\sim2\sigma$ deviations are mostly originated in the centres of voids and superclusters ($R/R_\mathrm{v,sc}<1$). The combination of voids and superclusters yielded a $10.3\sigma$ detection with an $A_{\kappa}\approx0.82\pm0.08$ constraint on the CMB lensing amplitude, i.e. the overall DES Y3 $\times$ \emph{Planck} $\kappa$ signal is $2.3\sigma$ weaker than expected from MICE.

We then also tested the role of different subsets of voids and superclusters in the low CMB lensing signals. We found that both voids and superclusters feature more discrepancy in the low-$z$ part of the DES Y3 data at redshifts $0.15<z<0.55$, with $A_{\kappa}\approx0.55\pm0.23$ and $A_{\kappa}\approx0.70\pm0.15$, respectively. While the internal consistency of the supercluster sample is generally better, we also observed that deeper voids ($\delta_{\rm c}<-0.6$ central under-density) imprint a lensing signal that is more anomalous with $A_{\kappa}\approx0.56\pm0.14$. Similarly, larger voids ($R_{\rm v}>35~\mpc$) also imprint a rather weak CMB $\kappa$ signal with an $A_{\kappa}\approx0.66\pm0.15$ amplitude. 

\begin{figure}
\begin{center}
\includegraphics[width=87mm
]{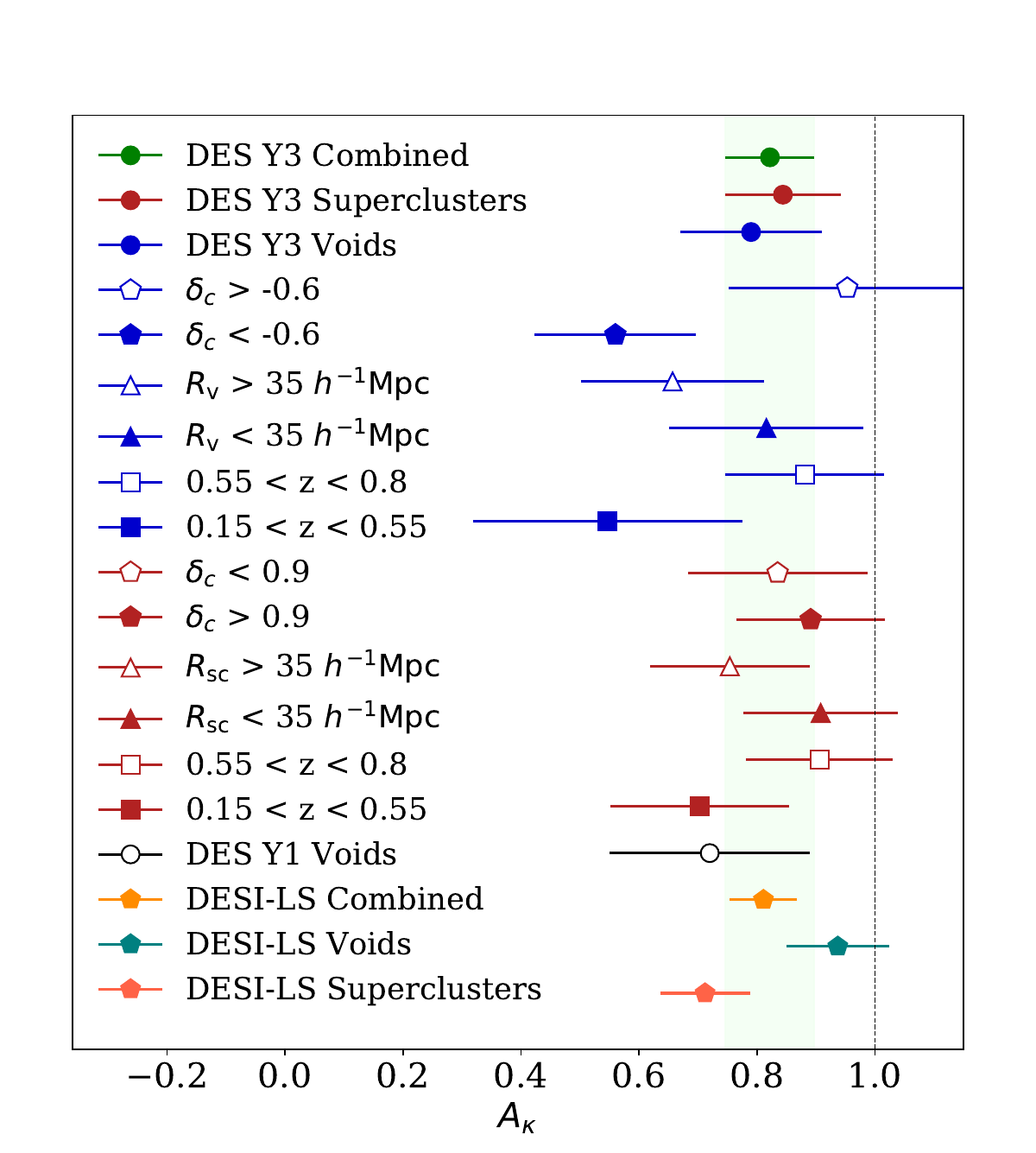}
\end{center}
\caption{Void, supercluster, and combined constraints on the $A_{\kappa}$ parameter, including the outcomes from the measurements using bins of central under-density, radius, and redshift. Moderately significant trends are seen for weaker signals from various subsets of the void data, while the supercluster bins are more consistent with each other. We also compare our DES Y3 results with the constraints from the DESI Legacy \emph{Imaging} Survey, and from voids detected in the smaller DES Y1 data set.}
\label{fig:profile_likelihoods_comp}
\end{figure}

As summarised in Figures \ref{fig:profile_likelihoods_comp}, \ref{fig:profile_voidsub}, and \ref{fig:profile_scsub}, the observed tensions from subsets of voids and superclusters are not highly significant, but they serve as valuable additions to the fiducial analysis with all objects used in the CMB $\kappa$ stacking measurements.

Yet, it is certainly interesting to note that some of the recent CMB $\kappa$ measurements from voids and superclusters also found lower-than-expected amplitudes. Using voids detected in the DES Y1 data set, \cite{Vielzeuf2019} reported $A_{\kappa}\approx0.72\pm0.17$ based on a similar \redmagic\ galaxy sample from about one third of the area that we analysed in this paper. While the significance of this DES Y1 result was low, and thus no significant tension was reported, \cite{Hang2021} analysed a larger sample of similarly defined voids and superclusters using the DESI-LS photo-$z$ catalogue, and reported $A_{\kappa}\approx0.811\pm0.057$, i.e. $3.3\sigma$ lower signal than expected from a standard $\Lambda$CDM model ($A_{\rm \kappa}\approx0.937\pm0.087$ for voids alone, and $A_{\rm \kappa}\approx0.712\pm0.076$ for superclusters). 

As in the case of the ISW excess signals, we note that not all void samples show anomalously low lensing signals \citep[see e.g.][]{Cai2017,Raghunathan2019}, and therefore more work is needed to settle this debate. For more solid conclusions, the cosmology-dependence of the CMB $\kappa$ signal from voids and superclusters should also be understood in greater detail.

Nonetheless, it is plausible to contemplate that the lower-than-expected lensing signals from low-$z$ voids and superclusters are due to a genuine physical effect (not excluding \emph{systematics}), since there are intriguing precedents for similar findings in cosmology:
\begin{itemize}
\item a strong (and unexplained) preference was observed for a lower linear galaxy bias parameter $b$ inferred from lensing compared to galaxy clustering ($X_{\rm lens}\approx0.87\pm0.02$) in the main $3\times2$-pt cosmological analysis of the DES Y3 data set in general \citep[][]{descollaboration2021dark}, and in the clustering plus galaxy-galaxy lensing analyses of the DES Y3 \redmagic\ sample in particular \citep[see][for details]{Pandey2021}.
\item the BOSS $\times$ CFHTLenS galaxy-galaxy lensing measurements by \cite{Leauthaud2017} showed that the observed lensing signal is $\sim$20-40$\%$ lower than expected based on the auto-correlation of the galaxy sample. Along similar lines, \cite{Lange2021} found that such tensions do not significantly depend on the halo mass in the $10^{13.3}$-$10^{13.9}h^{-1}M_\odot$ range (i.e. similar to \redmagic\ host halos), and no significant scale-dependence was observed in the $0.1\mpc<r<60\mpc$ range. These interesting results can exclude some proposed small-scale phenomena as explanations, such as baryonic effects or insufficient halo occupation models.
\item beyond the low amplitude from voids and superclusters with $A_{\rm \kappa}\approx0.811\pm0.057$ \citep[][]{Hang2021}, the power spectrum analysis of the DESI-LS galaxies and the \emph{Planck} CMB $\kappa$ map reported $A_{\rm \kappa}\approx0.901\pm0.026$ \citep[][]{Hang20212pt}. Similar analyses using LRGs selected from the DESI imaging survey data at $z < 1$ \citep{Kitanidis2021,White2021} also prefer lower galaxy bias values (possibly due to stronger dark matter clustering) and/or a lower lensing amplitude ($S_{8}=\sigma_{8}\sqrt{\Omega_{m}/0.3}\approx0.73\pm0.03$) than expected from a baseline \emph{Planck} 2018 cosmology ($S_{8}\approx0.832\pm0.012$).
\end{itemize}

\section{Conclusions}
\label{sec:Section5}

While remnant systematic effects remain plausible culprits for such apparent tensions, a possible interpretation of these findings, including the DES Y3 results of this paper, is that voids and superclusters show a more pronounced discrepancy because they probe the troughs and peaks of the density field more efficiently where the breakdown of the modelling approaches is most probable. On the other hand, it is also possible that the apparent tension is due to the limitations of the MICE simulation in such extreme environments.

Taken at face value, a lower-than-expected CMB lensing signal could, in principle, be a consequence of a faster low-$z$ expansion rate \citep[see e.g.][]{Riess2021} and a related stronger \emph{decay} of the gravitational potentials ($\dot{\Phi}<0$) than assumed in the baseline $\Lambda$CDM model. If the imbalance between cosmic expansion and structure growth is more pronounced (at least at the largest scales and in extreme density fluctuations) then the gravitational potential of these cosmic super-structures may evolve more strongly than in the standard model. As a consequence, their ability to deflect the paths of the CMB photons might also be reduced in comparison to baseline $\Lambda$CDM expectations, i.e. the too-low lensing and too-strong ISW signals are not necessarily inconsistent \citep[see][for a recent DES Y3 ISW analysis using supervoids]{Kovacs2019}. Future measurements of the CMB lensing and the ISW signals from voids and superclusters from the next generation of cosmological surveys (such as DESI, LSST, Euclid, or J-PAS) will provide decisive constraints on the validity of these interesting, yet not highly significant tensions.

\section*{Data availability}

Galaxy catalogues and masks are publicly available as part of the release of the DES Year-3 products\footnote{\url{https://www.darkenergysurvey.org/the-des-project/data-access/}}. The {\it Planck} CMB lensing map that we used to detect the imprints of voids and superclusters is also publicly available\footnote{\url{https://www.cosmos.esa.int/web/planck}}. The data underlying this article including the void catalogues, supercluster catalogues, and the measured lensing profiles will be shared on reasonable request to the corresponding author, in line with the relevant DES data policy.

\section*{Acknowledgments}

This paper has gone through \emph{internal review} by the DES collaboration. AK has been supported by a Juan de la Cierva \emph{Incorporaci\'on} fellowship with project number IJC2018-037730-I, and funding for this project was also available in part through SEV-2015-0548 and AYA2017-89891-P.

Funding for the DES Projects has been provided by the U.S. Department of Energy, the U.S. National Science Foundation, the Ministry of Science and Education of Spain, the Science and Technology Facilities Council of the United Kingdom, the Higher Education Funding Council for England, the National Center for Supercomputing Applications at the University of Illinois at Urbana-Champaign, the Kavli Institute of Cosmological Physics at the University of Chicago, the Center for Cosmology and Astro-Particle Physics at the Ohio State University, the Mitchell Institute for Fundamental Physics and Astronomy at Texas A\&M University, Financiadora de Estudos e Projetos, Funda{\c c}{\~a}o Carlos Chagas Filho de Amparo {\`a} Pesquisa do Estado do Rio de Janeiro, Conselho Nacional de Desenvolvimento Cient{\'i}fico e Tecnol{\'o}gico and the Minist{\'e}rio da Ci{\^e}ncia, Tecnologia e Inova{\c c}{\~a}o, the Deutsche Forschungsgemeinschaft and the Collaborating Institutions in the Dark Energy Survey. 

The Collaborating Institutions are Argonne National Laboratory, the University of California at Santa Cruz, the University of Cambridge, Centro de Investigaciones Energ{\'e}ticas, Medioambientales y Tecnol{\'o}gicas-Madrid, the University of Chicago, University College London, the DES-Brazil Consortium, the University of Edinburgh, the Eidgen{\"o}ssische Technische Hochschule (ETH) Z{\"u}rich, Fermi National Accelerator Laboratory, the University of Illinois at Urbana-Champaign, the Institut de Ci{\`e}ncies de l'Espai (IEEC/CSIC), the Institut de F{\'i}sica d'Altes Energies, Lawrence Berkeley National Laboratory, the Ludwig-Maximilians Universit{\"a}t M{\"u}nchen and the associated Excellence Cluster Universe, the University of Michigan, NFS's NOIRLab, the University of Nottingham, The Ohio State University, the University of Pennsylvania, the University of Portsmouth, SLAC National Accelerator Laboratory, Stanford University, the University of Sussex, Texas A\&M University, and the OzDES Membership Consortium.

Based in part on observations at Cerro Tololo Inter-American Observatory at NSF's NOIRLab (NOIRLab Prop. ID 2012B-0001; PI: J. Frieman), which is managed by the Association of Universities for Research in Astronomy (AURA) under a cooperative agreement with the National Science Foundation.

The DES data management system is supported by the National Science Foundation under Grant Numbers AST-1138766 and AST-1536171.
The DES participants from Spanish institutions are partially supported by MICINN under grants ESP2017-89838, PGC2018-094773, PGC2018-102021, SEV-2016-0588, SEV-2016-0597, and MDM-2015-0509, some of which include ERDF funds from the European Union. IFAE is partially funded by the CERCA program of the Generalitat de Catalunya.
Research leading to these results has received funding from the European Research Council under the European Union's Seventh Framework Program (FP7/2007-2013) including ERC grant agreements 240672, 291329, and 306478. We acknowledge support from the Brazilian Instituto Nacional de Ci\^encia e Tecnologia (INCT) do e-Universo (CNPq grant 465376/2014-2).

This manuscript has been authored by Fermi Research Alliance, LLC under Contract No. DE-AC02-07CH11359 with the U.S. Department of Energy, Office of Science, Office of High Energy Physics.

\bibliographystyle{mnras}
\bibliography{refs}

\section*{Author Affiliations}
{\small
$^{1}$ Instituto de Astrof\'{\i}sica de Canarias (IAC), Calle V\'{\i}a L\'{a}ctea, E-38200, La Laguna, Tenerife, Spain\\
$^{2}$ Departamento de Astrof\'{\i}sica, Universidad de La Laguna (ULL), E-38206, La Laguna, Tenerife, Spain\\
$^{3}$ Aix-Marseille Univ, CNRS/IN2P3, CPPM, Marseille, France\\
$^{4}$ Institute of Theoretical Astrophysics, University of Oslo. P.O. Box 1029 Blindern, NO-0315 Oslo, Norway\\
$^{5}$ Institut d'Estudis Espacials de Catalunya (IEEC), 08034 Barcelona, Spain\\
$^{6}$ Institute of Space Sciences (ICE, CSIC),  Campus UAB, Carrer de Can Magrans, s/n,  08193 Barcelona, Spain\\
$^{7}$ Institut de F\'{\i}sica d'Altes Energies (IFAE), The Barcelona Institute of Science and Technology, Campus UAB, 08193, Bellaterra, Spain\\
$^{8}$ Instituci\'o Catalana de Recerca i Estudis Avan\c{c}ats, Barcelona, Spain\\
$^{9}$ Department of Astronomy and Astrophysics, University of Chicago, Chicago, IL 60637, USA\\
$^{10}$ Kavli Institute for Cosmological Physics, University of Chicago, Chicago, IL 60637, USA\\
$^{11}$ Universit\"ats-Sternwarte, Fakult\"at f\"ur Physik, Ludwig-Maximilians Universit\"at M\"unchen, Scheinerstr. 1, 81679 M\"unchen, Germany\\
$^{12}$ Physics Department, 2320 Chamberlin Hall, University of Wisconsin-Madison, 1150 University Avenue Madison, WI  53706-1390\\
$^{13}$ Argonne National Laboratory, 9700 South Cass Avenue, Lemont, IL 60439, USA\\
$^{14}$ Center for Astrophysical Surveys, National Center for Supercomputing Applications, 1205 West Clark St., Urbana, IL 61801, USA\\
$^{15}$ Department of Astronomy, University of Illinois at Urbana-Champaign, 1002 W. Green Street, Urbana, IL 61801, USA\\
$^{16}$ Physics Department, William Jewell College, Liberty, MO, 64068\\
$^{17}$ Fermi National Accelerator Laboratory, P. O. Box 500, Batavia, IL 60510, USA\\
$^{18}$ Center for Cosmology and Astro-Particle Physics, The Ohio State University, Columbus, OH 43210, USA\\
$^{19}$ Department of Physics, The Ohio State University, Columbus, OH 43210, USA\\
$^{20}$ Department of Physics and Astronomy, University of Pennsylvania, Philadelphia, PA 19104, USA\\
$^{21}$ Kavli Institute for Cosmological Physics, University of Chicago, Chicago, IL 60637, USA\\
$^{22}$ SLAC National Accelerator Laboratory, Menlo Park, CA 94025, USA\\
$^{23}$ Centro de Investigaciones Energ\'eticas, Medioambientales y Tecnol\'ogicas (CIEMAT), Madrid, Spain\\
$^{24}$ Brookhaven National Laboratory, Bldg 510, Upton, NY 11973, USA\\
$^{25}$ Cerro Tololo Inter-American Observatory, NSF's National Optical-Infrared Astronomy Research Laboratory, Casilla 603, La Serena, Chile\\
$^{26}$ Laborat\'orio Interinstitucional de e-Astronomia - LIneA, Rua Gal. Jos\'e Cristino 77, Rio de Janeiro, RJ - 20921-400, Brazil\\
$^{27}$ Departamento de F\'isica Matem\'atica, Instituto de F\'isica, Universidade de S\~ao Paulo, CP 66318, S\~ao Paulo, SP, 05314-970, Brazil\\
$^{28}$ Institute of Cosmology and Gravitation, University of Portsmouth, Portsmouth, PO1 3FX, UK\\
$^{29}$ CNRS, UMR 7095, Institut d'Astrophysique de Paris, F-75014, France\\
$^{30}$ Sorbonne Universit\'es, UPMC Univ Paris 06, UMR 7095, Institut d'Astrophysique de Paris, F-75014, Paris, France\\
$^{31}$ University Observatory, Faculty of Physics, Ludwig-Maximilians-Universit\"at, Scheinerstr. 1, 81679 Munich, Germany\\
$^{32}$ Department of Physics \& Astronomy, University College London, Gower Street, London, WC1E 6BT, UK\\
$^{33}$ Astronomy Unit, Department of Physics, University of Trieste, via Tiepolo 11, I-34131 Trieste, Italy\\
$^{34}$ INAF-Osservatorio Astronomico di Trieste, via G. B. Tiepolo 11, I-34143 Trieste, Italy\\
$^{35}$ Institute for Fundamental Physics of the Universe, Via Beirut 2, 34014 Trieste, Italy\\
$^{36}$ Observat\'orio Nacional, Rua Gal. Jos\'e Cristino 77, Rio de Janeiro, RJ - 20921-400, Brazil\\
$^{37}$ Department of Physics, University of Michigan, Ann Arbor, MI 48109, USA\\
$^{38}$ Department of Physics, IIT Hyderabad, Kandi, Telangana 502285, India\\
$^{39}$ Universit\"ats-Sternwarte, Fakult\"at f\"ur Physik, Ludwig-Maximilians Universit\"at M\"unchen, Scheinerstr. 1, 81679 M\"unchen, Germany\\
$^{40}$ Jet Propulsion Laboratory, California Institute of Technology, 4800 Oak Grove Dr., Pasadena, CA 91109, USA\\
$^{41}$ Instituto de Fisica Teorica UAM/CSIC, Universidad Autonoma de Madrid, 28049 Madrid, Spain\\
$^{42}$ Department of Astronomy, University of Michigan, Ann Arbor, MI 48109, USA\\
$^{43}$ Institute of Astronomy, University of Cambridge, Madingley Road, Cambridge CB3 0HA, UK\\
$^{44}$ Kavli Institute for Cosmology, University of Cambridge, Madingley Road, Cambridge CB3 0HA, UK\\
$^{45}$ School of Mathematics and Physics, University of Queensland,  Brisbane, QLD 4072, Australia\\
$^{46}$ Santa Cruz Institute for Particle Physics, Santa Cruz, CA 95064, USA\\
$^{47}$ Australian Astronomical Optics, Macquarie University, North Ryde, NSW 2113, Australia\\
$^{48}$ Lowell Observatory, 1400 Mars Hill Rd, Flagstaff, AZ 86001, USA\\
$^{49}$ George P. and Cynthia Woods Mitchell Institute for Fundamental Physics and Astronomy, and Department of Physics and Astronomy, Texas A\&M University, College Station, TX 77843,  USA\\
$^{50}$ Department of Astrophysical Sciences, Princeton University, Peyton Hall, Princeton, NJ 08544, USA\\
$^{51}$ Perimeter Institute for Theoretical Physics, 31 Caroline St N, Waterloo, Canada\\
$^{52}$ Department of Astronomy, University of California, Berkeley,  501 Campbell Hall, Berkeley, CA 94720, USA\\
$^{53}$ School of Physics and Astronomy, University of Southampton,  Southampton, SO17 1BJ, UK\\
$^{54}$ Computer Science and Mathematics Division, Oak Ridge National Laboratory, Oak Ridge, TN 37831\\
$^{55}$ Excellence Cluster Origins, Boltzmannstr.\ 2, 85748 Garching, Germany\\
$^{56}$ Max Planck Institute for Extraterrestrial Physics, Giessenbachstrasse, 85748 Garching, Germany\\}
\end{document}